\documentclass[12pt]{article}

\RequirePackage[OT1]{fontenc}
\RequirePackage{amsthm,amsmath}
\RequirePackage[round, authoryear]{natbib}
\RequirePackage[colorlinks,linkcolor=blue,citecolor=blue,urlcolor=blue]{hyperref}
\usepackage{times}
\usepackage{bm}
\usepackage{amsmath}
\usepackage{amsfonts}
\usepackage{amsthm} 
\usepackage{amssymb}
\usepackage{amstext}
\usepackage{bbm}
\usepackage{mathtools}
\usepackage{graphicx}
\usepackage{adjustbox}
\graphicspath{ {./figures/} }
\usepackage[mathscr]{euscript}
\usepackage{tabularx}
\usepackage{mathrsfs}
\usepackage{enumerate} 
\usepackage{cases} 
\usepackage{float} 
\usepackage{caption}
\usepackage{subcaption}
\usepackage{comment}
\usepackage{wrapfig}
\usepackage[english]{babel}
\usepackage[utf8]{inputenc}
\usepackage[colorinlistoftodos]{todonotes}
\usepackage[T1]{fontenc}
\usepackage{amsbsy} 
\usepackage{physics} 
\usepackage{upgreek} 
\usepackage{diagbox} 
\usepackage{rotating}
\usepackage{authblk}
\usepackage{booktabs}
\usepackage{multirow}
\usepackage[normalem]{ulem}

\theoremstyle{plain}
\newtheorem{definition}{Definition}[section]
\newtheorem{theorem}{Theorem}[section]
\newtheorem{lemma}[theorem]{Lemma}

\newtheorem{cor}{Corollary}[section]

\theoremstyle{definition}

\theoremstyle{remark}

\def\bbE{{\mathbbm E}}
\def\Var{{\mathrm {Var}}}
\def\Cov{{\mathrm {Cov}}}

\def\blue{\color{blue}}

\def \bsm {\bm}
\def \btheta {\bsm{\theta}}

\def \bbeta {\bsm{\beta}}

\def \bdeta {\bsm{\eta}}

\def \bone {\bsm{\mathrm{1}}} 
\def \bq {\mathbf{q}}

\def \ba {\boldsymbol{a}}

\def \by {\boldsymbol{y}}

\def \bQ {\mathbf{Q}}

\def \bI {\mathbf{I}}

\DeclareMathOperator*{\argmax}{arg\,max} 
\newcommand{\ind}{\mbox{$\mathbbm{1}$}}

\def \bea {\begin{eqnarray*}}
\def \eea {\end{eqnarray*}}

\title{Optimal Estimation under a Semiparametric Density Ratio Model}
\author[1]{Archer Gong Zhang \thanks{Corresponding author. Email address: \url{archer.zhang@utoronto.ca}}}
\author[2]{Jiahua Chen}
\affil[1]{Department of Statistical Sciences, University of Toronto, Toronto, ON, Canada, M5G 1Z5} 
\affil[2]{Department of Statistics, University of British Columbia, Vancouver, BC, Canada, V6T 1Z4} 
\date{}

\begin{document}
\maketitle

\begin{abstract} 
In many statistical and econometric applications, we gather individual samples from various interconnected populations that undeniably exhibit common latent structures.
Utilizing a model that incorporates these latent structures for such data enhances the efficiency of inferences. 
Recently, many researchers have been adopting the semiparametric density ratio model (DRM) to address the presence of latent structures.
The DRM enables estimation of each population distribution using pooled data, resulting in statistically more efficient estimations in contrast to nonparametric methods that analyze each sample in isolation.
In this article, we investigate the limit of the efficiency improvement attainable through the DRM. 
We focus on situations where one population's sample size significantly exceeds those of the other populations. 
In such scenarios, we demonstrate that the DRM-based inferences for populations with smaller sample sizes achieve the highest attainable asymptotic efficiency as if a parametric model is assumed. 
The estimands we consider include the model parameters, distribution functions, and quantiles. 
We use simulation experiments to support the theoretical findings with a specific focus on quantile estimation. 
Additionally, we provide an analysis of real revenue data from U.S. collegiate sports to illustrate the efficacy of our contribution.
\end{abstract}


\section{Introduction} 
\label{sec:intro} 

In numerous applications, scientists may independently collect a single sample from each of the connected and similar populations. 
The dataset hence contains independent multiple samples. 
For example, to gauge the economic state of a country, econometricians may study the evolution of income distribution over time \citep{roine2015long}. 
They may hence collect cross-sectional or panel survey data from year to year \citep{wooldridge2010econometric}, and the underlying populations for these multiple samples are naturally connected. 
There are various approaches to data analysis of this nature. 
One may postulate a parametric model such as normal or gamma on these populations \citep{davison2003statistical}. 
However, some statistical inference procedures such as quantile estimation may have poor performance when the model assumptions are merely mildly violated. 
For instance, statistical analyses of a dataset under two similar models, log-normal and gamma models, can yield markedly different results \citep{wiens1999log}. 
To mitigate the risks of model misspecification, one may take a nonparametric approach \citep{wasserman2006all} by not imposing any parametric assumption. 
However, the nonparametric approaches ignore the latent structures shared by the multiple populations from which the samples are collected, and hence fail to utilize the potential gain in statistical efficiency. 
As a balanced trade-off between the model misspecification risks and the statistical efficiency, we advocate semiparametric approaches \citep{hardle2004nonparametric, tsiatis2006semiparametric}. 

Recently, there have been many research studies on the semiparametric density ratio model (DRM) \citep{anderson1979multivariate}. 
Let $ g_0 (x), \ldots, g_m (x) $ be the density functions for the multiple populations. 
The DRM postulates that 
\begin{align} 
g_k (x) = g_0 (x) \exp \{ \btheta_k^{\top} \bq (x) \}, 
\hspace{5mm} 
k = 1, \ldots, m, 
\label {DRM}
\end{align} 
where $ \bq (x) $ is a prespecified vector-valued function and $ \btheta_k $ is an unknown vector-valued parameter.
The DRM effectively utilizes the latent structures shared by the multiple populations. 
Not surprisingly, methods developed under the DRM are often statistically more efficient than the nonparametric methods that ignore the latent structures and use the samples separately for individual populations.
In particular, the DRM-based empirical likelihood (EL) approaches \citep{owen2001empirical} are found to have nice asymptotic properties and superior performances \citep{qin1993empirical, chen2013quantile, cai2017hypothesis}. 
Under the framework of the DRM and EL, the multiple samples are used together to draw inferences on each population. 

These results lead to an interesting question: what is the limit of the efficiency gain through the DRM-based approach? 
A gold-standard target to consider is the parametric efficiency. 
As a semiparametric model, a DRM naturally contains many parametric models, with each forming a so-called parametric submodel. 
For every such parametric submodel, there is a classical Cram\'{e}r--Rao lower bound. 
As a result, the asymptotic variance for any consistent and asymptotically normal semiparametric estimator is no smaller than the supremum of the Cram\'{e}r--Rao lower bounds for all parametric submodels. 
Such a supremum is famously known as the semiparametric efficiency bound \citep{newey1990semiparametric}. 
When a parametric submodel contains the true distributions that generate the data and is regular, the parametric maximum likelihood estimator (MLE) has the smallest asymptotic variance that attains the Cram\'{e}r--Rao lower bound \citep{casella2002statistical}. 
Therefore, no methods can achieve higher efficiency under the DRM than it can under its parametric submodel that the true distributions reside. 
This gives an efficiency upper bound for any semiparametric approach including the DRM-based estimators. 
Motivated by this observation, our research question becomes: is it likely that the DRM-based approaches achieve parametric efficiency under a parametric submodel? 
If yes, when would that happen?

In this article, we focus on the scenario where the sample size from one population is significantly larger than the sample sizes from the other populations.  
We are interested in the efficiency of the DRM-based estimators for the populations with smaller samples. 
In other words, we want to quantify how much help the large sample population can offer to improve the efficiency of the estimators for the small sample populations. 
The motivation for us to consider this scenario is as follows. 
Suppose the sample size from one population is infinite. 
The corresponding population distribution can therefore be regarded as known.
Consequently, treating the infinite-sample population density as $ g_0 (x) $, the semiparametric DRM in \eqref{DRM} would reduce to a parametric model for the other populations. 
We therefore expect that the DRM-based estimators for the small sample populations will achieve parametric efficiency under the corresponding parametric models. 
This article establishes and proves this claim for the model parameters, distribution functions, and quantiles.
Although we prove these results mostly in a two-sample scenario for clarity, our discoveries are general for the multi-sample scenario.
More specifically, when one population has an extra large sample, the DRM-based estimators for the other populations are as efficient as the best possible under the true parametric submodel. 
For convenience, we refer to this scenario as the two-sample scenario throughout the article. 

In the existing literature, there have been some discussions on the asymptotic efficiency of the DRM-based approaches, such as \citet{zhang2000quantile} and \citet{chen2013quantile}. 
They focus on situation where the sample sizes all go to infinity at the same rate and find that the DRM-based quantile estimators are asymptotically more efficient than the nonparametric empirical quantiles. 
However, none of these studies systematically investigate the boundary of the efficiency gain under the DRM. 
This article tackles this important research problem by examining the aforementioned two-sample scenario. 
Furthermore, their theoretical results rely on the assumption that the ratio of any two sample sizes remains fixed, positive, and finite as the sample sizes aproach infinity. 
In contrast, we study the asymptotic efficiency of the DRM-based approaches in a different and more general context that allows the sample sizes to grow at different rates and allows the ratios of sample sizes to evolve. 

The rest of the article is organized as follows. 
In Section~\ref{sec:DRM}, we elaborate more on the DRM and state the research problem. 
For comparison, in Section~\ref{sec:eff-submodel} we present the gold-standard parametric efficiency under a specific parametric submodel of the DRM. 
In Section~\ref {sec:EL-DRM}, we give an overview of the EL-based inferences under the DRM, and study the asymptotic efficiency of the EL-DRM estimators of the model parameters, distribution functions, and quantiles. 
For illustration, we show in Section~\ref{sec:DRM-identical} how the asymptotic variance of the DRM-based quantile estimator evolves as a function of the ratio of the sample sizes when the true population distributions $ G_k $ are all identical but not assumed to be identical in the estimation.  
We conduct some simulation studies in Section~\ref{sec:simulations} by generating data from normal and exponential distributions, whose results support our theoretical findings. 
In Section~\ref{sec:realdata}, we report results from a real-data analysis on the revenues of U.S. collegiate sports. 
Finally, Section~\ref{sec:conclusion} concludes our contributions in the article with a discussion. 
Proofs of the theoretical results are provided in \hyperref[sec:appendix]{Appendix}.

\section{The research problem} 
\label{sec:DRM} 

Suppose we have $ m+1 $ independent sets of independent and identically distributed (i.i.d.) samples respectively from population distributions $ G_0, \ldots, G_m $: 
\begin{align} 
x_{k, 1}, \ldots, x_{k, n_k} \overset {\text {i.i.d.}} \sim G_k, 
\hspace{5mm} 
k = 0, \ldots, m, 
\label{multi-sample}
\end{align} 
Let $ g_0 (\cdot), \ldots, g_m (\cdot) $ be the corresponding density functions with respect to some common $ \sigma $-finite measure. 
They satisfy the DRM if 
\begin{align*} 
g_k (x) = g_0 (x) \exp \{ \btheta_k^{\top} \bq (x) \}, 
\hspace{5mm} 
k = 1, \ldots, m, 
\end{align*} 
with $ \bq (x) $ a prespecified vector-valued function of dimension $ d $ 
and $ \btheta^{\top} \coloneqq (\btheta_1^{\top}, \ldots. \btheta_m^{\top}) $ an unknown vector-valued parameter. 
The first element of $ \bq (x) $ is set to be 1 so that the corresponding coefficient in $ \btheta_k $ is a normalization constant. 
Therefore, we may write $ \bq (x) $ as 
\begin{align}
\bq^{\top} (x) = (1, \bq_{-}^{\top} (x)), 
\label{q_}
\end{align}
for some vector-valued function $ \bq_{-}^{\top} (x) $ with dimension $ d-1 $. 
Correspondingly, we may also decompose the parameter $ \btheta_k $ into $ (\alpha_k, \bbeta_k) $. 
We also require the elements of $ \bq (x) $ to be linearly independent; otherwise, some elements of $ \bq (x) $ are redundant. 
By convention, we call $ G_0 $ the base distribution and $ \bq (x) $ the basis function. 
The DRM and its inference, which we introduce later, are invariant to the choice of the base distribution. 
In this article, we select the population distribution from which we have a much larger sample as the base distribution. 

The DRM covers many well-known parametric distribution families, with different choices of the base distribution $ G_0 $ and the basis function $ \bq (x) $. 
For example, 
the Poisson distribution family satisfies the DRM with $ G_0 $ being Poisson and $ \bq (x) = (1, x)^{\top} $; 
the normal distribution family satisfies the DRM with $ G_0 $ being normal and $ \bq (x) = (1, x, x^{2})^{\top} $; 
and the gamma distribution family also satisfies the DRM with $ G_0 $ being gamma and $ \bq (x) = (1, x, \log x)^{\top} $. 
In fact, any exponential family model satisfies the DRM. 
On the other hand, when using DRM, we do not assume that each population distribution belongs to any parametric distribution family. 
Instead of directly modelling the distribution of each population, the DRM takes a semiparametric approach to modelling the relationship between the multiple populations. 
Therefore, we may see that the DRM is a very flexible model and has a low risk of model misspecification. 
A more appealing feature of the DRM is that, with an appropriately chosen basis function $ \bq (x) $, the DRM allows us to utilize the combined data to estimate each distribution $ G_k $. 
This would lead to efficiency gain compared with estimating $ G_k $ individually using the sample from itself. 
The DRM has been widely applied in many fields; see the books \citet{sugiyama2012density} and \citet{qin2017biased}. 
It is also closely connected with other semiparametric models such as 
models for biased sampling \citep{vardi1982nonparametric, vardi1985empirical},
the exponential tilting model \citep{rathouz2009generalized},
and the proportional likelihood ratio model \citep{luo2012proportional}.
Many studies on a variety of inference problems in the literature have confirmed this efficiency gain through either theoretical or empirical analyses, including \citet{zhang2000quantile, chen2013quantile, cai2017hypothesis}.

The research problem we consider in this article is: when the sample from $ G_0 $ has a much larger size than the samples from the other $ G_k $, would the DRM-based estimators of some functionals of $ G_k $ achieve parametric efficiency under a parametric submodel? 
Without loss of generality and for clarity, hereafter we present our results mostly in the two-sample scenario, namely $ m = 1 $. 
Our results are general for the multiple samples as long as we have a much larger sample from one population than the others. 
We denote by $ n_k $ the size of the sample from $ G_k $, and let $ n = \sum_{k=0}^{m} n_k $ be the total sample size. 
Formally, when $ G_0, G_1 $ satisfy the DRM, we study the asymptotic efficiency of the estimators of some functionals of $ G_1 $ when 
\begin{align}
n_0/n_1 \to \infty \, \, \, \text{ as } \, \, \, n_0, n_1 \to \infty. 
\label{two-sample-scenario}
\end{align}
The functionals of $ G_1 $ we consider include the model parameters $ \btheta $, the cumulative distribution function $ G_1 (x) $ for every $ x $, and the quantiles of $ G_1 $. 

The research problem has many applications in statistics and econometrics. 
For example, an important aspect of econometrics is to analyze cross-sectional and panel data with some structures \citep{ng2006testing, chen2012semiparametric, lin2012estimation, boneva2015semiparametric, su2016identifying, gao2020heterogeneous, su2023identifying}, which is useful for areas of economic research such as understanding the economic state of a population at specific time points and evaluating the effects of a policy across different groups, regions, or time periods. 
The DRM has been found successful for the analysis of such data in the long-term monitoring of lumber strength properties \citep{zidek2018statistical, chen2022permutation}. 
This is because the DRM-based methods are able to borrow strength across the multiple cross-sectional samples to exploit the shared latent structures between different populations over years and regions. 
Therefore, it is of scientific significance to quantify the resulting efficiency gain when analyzing such type of data. 
Our contributions in this article achieve this aim by focusing on the situation where one wishes to make inferences on a population with a small data set, aided by a large data set from another connected population. 
In the real-data analysis section of this article, we study the efficiency of the DRM-based quantile estimators under such scenario based on a revenue data from U.S. collegiate sports.

\section{Estimation efficiency under a parametric submodel}
\label{sec:eff-submodel} 

To study the efficiency of the DRM-based inferences, we first reveal the efficiency limit under its parametric submodels. 
A parametric submodel is a parametric family of distributions we select for the data that satisfy the semiparametric model (in our case, the DRM) assumptions. 
In this article, the parametric submodel we consider for $ G_0, G_1 $ is an exponential family model: 
\begin{align}
g_0 (x) & = B (x) \exp \{ \bdeta_0^{\top} \bq_{-} (x) + A (\bdeta_0) \}, \nonumber \\ 
g_1 (x) & = B (x) \exp \{ \bdeta_1^{\top} \bq_{-} (x) + A (\bdeta_1) \}, \label{submodel}
\end{align}
where $ \bdeta_0, \bdeta_1 $ are the unknown natural parameters to be estimated, 
$ \bq_{-} (x) $ is the sufficient statistic for a single observation $ x $ that is the same as the one in \eqref{q_} in the DRM, 
$ B (\cdot) $ is a known nonnegative function, 
and $ A (\bdeta_k) $ is uniquely determined by $ \bdeta_k $: 
\[ 
A (\bdeta_k) = -\log \int B (x) \exp \{ \bdeta_k^{\top} \bq_{-} (x) \} \mathrm{d} x, 
\] 
such that $ g_k $ is indeed a density function for $ k = 0, 1 $. 

{\bf Remark}: the results presented in the article are still valid if we consider another set of parametric submodel: $ g_0 (x) $ being completely known and $ g_1 (x) $ from an exponential family model. 
This submodel consists of joint distributions of two independent set of i.i.d. samples in the form of
\(
\prod_{i=1}^{n_0} g_0 (x_{0i}) 
\prod_{j=1}^{n_1} g_0 (x_{1j}) \exp \{ \alpha + \bbeta^{\top} \bq_{-} (x_{1j} ) \}. 
\)
 
With the decomposition of $ \bq (x) $ and $ \btheta $ in \eqref{q_}, we get an equivalent form of the DRM in \eqref{DRM}: 
\begin{align}
g_1(x; \btheta) 
= g_0(x) \exp\{ \btheta^\top \bq (x)\}
= g_0 (x) \exp \{ \alpha + \bbeta^\top \bq_{-} (x) \}, 
\label{DRM-decomp}
\end{align}
where $ \alpha $ is a normalization constant determined by 
\[
\alpha = -\log \int g_0(x) \exp \{ \bbeta^\top \bq_{-} (x) \} \mathrm{d} x.
\]
Therefore, the exponential family model in \eqref{submodel} satisfies the DRM in \eqref{DRM-decomp} with 
\[ 
\btheta = 
\begin{pmatrix}
\alpha \\ 
\bbeta
\end{pmatrix} 
= 
\begin{pmatrix}
A (\bdeta_1) - A (\bdeta_0) \\ 
\bdeta_1 - \bdeta_0
\end{pmatrix}
.
\] 
In other words, the exponential family model in \eqref{submodel} is indeed a parametric submodel of the DRM. 
In this section, we obtain the asymptotic results for $ G_1 $ under this parametric submodel.

\subsection{Estimation of \texorpdfstring{$\boldsymbol{\theta}$}{} under the parametric submodel}

We first study the efficiency of the parametric MLE of $ \btheta $ under the parametric submodel \eqref{submodel}. 
This will be our gold-standard target with which we will compare the efficiency of the DRM estimator of $ \btheta $ to be presented later. 
We obtain a parametric estimator of $ \btheta $ using the maximum likelihood method. 
The parametric log-likelihood function given the two independent i.i.d. samples is 
\begin{align*}
\ell_{\mathrm{para}} (\bdeta_0, \bdeta_1) 
& = \sum_{j=1}^{n_0} \log g_0 (x_{0, j}) + \sum_{j=1}^{n_1} \log g_1 (x_{1, j}) \\ 
& = n_0 A (\bdeta_0) + \bdeta_0^{\top} \sum_{j=1}^{n_0} \bq_{-} (x_{0, j}) + n_1 A (\bdeta_1) + \bdeta_1^{\top} \sum_{j=1}^{n_1} \bq_{-} (x_{1, j}) + \sum_{k, j} B (x_{k j}). 
\end{align*}
From this, we get the score function: for $ k = 0, 1 $, 
\[
\frac{\partial \ell_{\mathrm{para}} (\bdeta_0, \bdeta_1)}{\partial \bdeta_k} 
= n_k \frac{\partial A (\bdeta_k)} {\partial \bdeta_k} + \sum_{j=1}^{n_k} \bq_{-} (x_{k j}). 
\] 
We let $ \tilde \bdeta_k $ denote the parametric MLE of $ \bdeta_k $, and they satisfy  
\begin{align*}
\frac{\partial A (\tilde \bdeta_k)} {\partial \bdeta_k} = -n_k^{-1} \sum_{j=1}^{n_k} \bq_{-} (x_{k j}).
\end{align*} 
We remark that for convenience, with a generic function $ f (\by) $, 
we have used the notation 
\[ 
\frac{\partial f (\by^{*}) }{\partial \by} 
= 
\left. \frac{\partial f (\by) }{\partial \by} 
\right \rvert_{\by = \by^{*}} 
\] 
in the preceding expression, and we retain the use of such notation hereafter.  

By the invariance property of maximum likelihood estimation (see Theorem~\ref {invariance_MLE} below), which states that any function of MLE is still an MLE of the function of the parameter, 
we must have that the MLE of $ \btheta $, denoted as $ \tilde \btheta $, is given by 
\begin{align}
\label{theta-MLE}
\tilde \btheta 
= 
\begin{pmatrix}
A (\tilde \bdeta_1) - A (\tilde \bdeta_0) \\ 
\tilde \bdeta_1 - \tilde \bdeta_0
\end{pmatrix}
.
\end{align}
Before we formally state the invariance property of the MLEs, we first introduce the so-called induced likelihood \citep{casella2002statistical}, which is also known as the profile likelihood in the literature \citep{barndorff1988parametric}. 
Suppose a distribution family is indexed by a parameter $ \theta $, and we want to find the MLE of some function of $ \theta $, denoted by $ \tau (\theta) $. 
When $ \tau (\cdot) $ is one-to-one, then given the MLE of $ \theta $ being $ \hat \theta $, it can be seen that the MLE of $ \tau (\theta) $ is $ \tau (\hat \theta) $. 
When $ \tau (\cdot) $ is not one-to-one, there are some technical issues under the current definition of likelihood. 
To overcome these problems, we need a more general definition of the likelihood function of $ \tau (\theta) $, called the induced likelihood function, and correspondingly a more general definition of the MLE. 
They are defined as follows. 

\begin{definition}[Extended definitions of likelihood and MLE \citep {casella2002statistical}] 

Denote by $ L (\theta) $ the likelihood function of $ \theta $. 
The induced likelihood function for $ \eta = \tau (\theta) $ is defined as 
\begin{align*}
L^{*} (\eta) = \sup_{\{ \theta: \tau (\theta) = \eta \}} L (\theta). 
%\label {induced_likelihood}
\end{align*} 
Further, $ \hat \eta $ that maximizes the induced likelihood $ L^{*} (\eta) $ is defined as the MLE of $ \eta = \tau (\theta) $. 
 
\end{definition} 

With this extended definition of the MLE, we are now ready to formally state the invariance property of MLEs in the following theorem. 

\begin{theorem}[Invariance property of MLE \citep {casella2002statistical}] 
\label{invariance_MLE} 

If $ \tilde \theta $ is the MLE of $ \theta $, then for any function $ \tau (\theta) $, 
$ \tau (\tilde \theta) $ is an MLE of $ \tau (\theta) $. 

\end{theorem} 

After obtaining the parametric MLE $ \tilde \btheta $, we now study its asymptotic properties. 
By the standard results on maximum likelihood estimation theory, under the exponential family model in \eqref{submodel} and suppose some regularity conditions are satisfied, 
the MLE $ \tilde \bdeta_k $ is asymptotically normal: 
\begin{align}
\sqrt {n_k} (\tilde \bdeta_k - \bdeta_k^{*}) 
\overset {d} \to 
N (\bsm {0}, \Var_k^{-1} [\bq_{-} (X)]), 
\label{mle-eta-normal}
\end{align}
as $ n_k \to \infty $, 
where $ \bdeta_k^{*} $ denotes the true value of $ \bdeta_k $. 
Applying the delta method \citep {casella2002statistical}, we have that $ (A (\tilde \bdeta_k), \tilde \bdeta_k) $ is also asymptotically normal: 
\[ 
\sqrt {n_k} 
\begin{pmatrix}
A (\tilde \bdeta_k) - A (\bdeta_k^{*}) \\ 
\tilde \bdeta_k - \bdeta_k^{*}
\end{pmatrix}
\overset {d} \to 
N 
\left (
\bsm {0}, 
\begin{pmatrix}
- \bbE_k [\bq_{-}^{\top} (X)] \\ 
\bI_d
\end{pmatrix}
\Var_k^{-1} [\bq_{-} (X)]
\begin{pmatrix}
- \bbE_k [\bq_{-} (X)], 
\, 
\bI_d
\end{pmatrix}
\right ), 
\] 
as $ n_k \to \infty $. 
Note that here we have used a standard result on the exponential family model: 
\begin{align} 
\frac {\partial A (\bdeta_1^{*})} {\partial \bdeta_1} 
= - \bbE_1 [\bq_{-} (X)]. 
\label {A_derivative}
\end{align} 
Taking advantage of the nice mathematical properties of exponential families, we forego other otherwise necessary verifications.

Finally, because the two samples are independent of each other, 
we have 
\begin{align} 
\sqrt {n_1} (\tilde \btheta - \btheta^{*}) 
& = 
(n_1/n_0)^{1/2} 
\sqrt {n_0}
\begin{pmatrix}
A (\tilde \bdeta_1) - A (\bdeta_1^{*}) \\ 
\tilde \bdeta_1 - \bdeta_1^{*}
\end{pmatrix}
- 
\sqrt {n_1}
\begin{pmatrix}
A (\tilde \bdeta_0) - A (\bdeta_0^{*}) \\ 
\tilde \bdeta_0 - \bdeta_0^{*}
\end{pmatrix} 
\nonumber \\ 
& \overset {d} \to 
N 
\left (
\bsm {0}, 
\begin{pmatrix}
- \bbE_1 [\bq_{-}^{\top} (X)] \\ 
\bI_d
\end{pmatrix}
\Var_1^{-1} [\bq_{-} (X)]
\begin{pmatrix}
- \bbE_1 [\bq_{-} (X)], 
\, 
\bI_d
\end{pmatrix}
\right ),  
\label{mle-efficiency}
\end{align} 
where $ n_1/n_0 \to 0 $, as $ n_0, n_1 \to \infty $.  
We will compare the asymptotic variance in \eqref{mle-efficiency} of the parametric MLE $ \tilde \btheta $ with the asymptotic variance of the DRM-based estimator of $ \btheta $ later.

\subsection{Distribution estimation under the parametric submodel}
\label{sec:dist_Para} 

We next study the inference on the distribution function $ G_1 (x) $ under the two-sample exponential family model in \eqref{submodel}. 
By Theorem~\ref {invariance_MLE}, the MLE of $ G_1 (x) = G_1 (x; \bdeta_1^{*}) $ is given by 
\[ 
\tilde G_1 (x) = G_1 (x; \tilde \bdeta_1)
= \int_{-\infty}^{x} B (t) \exp \{ \tilde \bdeta_1^{\top} \bq_{-} (t) + A (\tilde \bdeta_1) \} \mathrm {d} t,
\] 
where we recall that $ \tilde \bdeta_1 $ is the MLE of $ \bdeta_1 $.

The limiting distribution of such a parametric distribution estimator $ \tilde G_1 (x) $ can be straightforwardly obtained, as follows. 
Recall that the MLE $ \tilde \bdeta_1 $ is asymptotically normal: 
\[ 
\sqrt {n_1} (\tilde \bdeta_1 - \bdeta_1^{*}) 
\overset {d} \to 
N (\bsm {0}, \Var_1^{-1} [\bq_{-} (X)]), 
\] 
as $ n_1 \to \infty $. 
Applying the delta method, we have that for every fixed $ x $, 
the MLE $ \tilde G_1 (x) $ is also asymptotically normal: 
\begin{align}
\sqrt {n_1} 
\{ \tilde G_1 (x) - G_1 (x) \} 
& = 
\sqrt {n_1} 
\{ G_1 (x; \tilde \bdeta_1) - G_1 (x; \bdeta_1^{*}) \} 
\nonumber \\ 
& \overset {d} \to 
N 
\left (
\bsm {0}, 
\frac {\partial G_1 (x; \bdeta_1^{*})} {\partial \bdeta_1^{\top}}
\Var_1^{-1} [\bq_{-} (X)]
\frac {\partial G_1 (x; \bdeta_1^{*})} {\partial \bdeta_1}
\right ), 
\hspace{2mm}
\text { as } n_1 \to \infty. 
\label {mle-dist-efficiency}
\end{align}

Under regularity conditions, we can further simplify the variance matrix in \eqref {mle-dist-efficiency} as follows. 
First, we note that 
\begin{align} 
\frac {\partial G_1 (x; \bdeta_1^{*})} {\partial \bdeta_1} 
& = \int_{-\infty}^{x} 
\left . 
\frac {\partial \exp \{ \bdeta_1^{\top} \bq_{-} (t) + A (\bdeta_1) \}} {\partial \bdeta_1} 
\right \vert_{\bdeta_1 = \bdeta_1^*}
B (t) \mathrm {d} t 
\nonumber \\ 
& = \int_{-\infty}^{x} 
\left \{ \bq_{-} (t) + \frac {\partial A (\bdeta_1^*)} {\partial \bdeta_1} \right \}
\exp \{ \bdeta_1^{* \top} \bq_{-} (t) + A (\bdeta_1^*) \} B (t) \mathrm {d} t 
\nonumber \\ 
& = \int_{-\infty}^{x} \bq_{-} (t) \mathrm {d} G_1 (t) 
+ \frac {\partial A (\bdeta_1^*)} {\partial \bdeta_1} \int_{-\infty}^{x} \mathrm {d} G_1 (t) 
\nonumber \\ 
& = \bQ (x) + G_1 (x) \frac {\partial A (\bdeta_1^{*})} {\partial \bdeta_1}, 
\label {G_derivative}
\end{align} 
by the definition of $ \bQ (x) $ given in Theorem~\ref {thm:dist_normality}. 
Further, recall from \eqref {A_derivative} that 
$
\partial A (\bdeta_1^{*}) / \partial \bdeta_1 
= - \bbE_1 [\bq_{-} (X)]. 
$
Therefore, the preparation results in \eqref {A_derivative} and \eqref {G_derivative} together lead to a simpler expression for the variance matrix in \eqref {mle-dist-efficiency}:  
\begin{align} 
& \frac {\partial G_1 (x; \bdeta_1^{*})} {\partial \bdeta_1^{\top}}
\Var_1^{-1} [\bq_{-}]
\frac {\partial G_1 (x; \bdeta_1^{*})} {\partial \bdeta_1} 
\nonumber \\ 
& = 
\left \{ \bQ (x) - \bbE_1 [\bq_{-}] G_1 (x) \right \}^{\top}
\Var_1^{-1} [\bq_{-}]
\left \{ \bQ (x) - \bbE_1 [\bq_{-}] G_1 (x) \right \}. 
\label{mle-dist-efficiency-eqv}
\end{align} 
We will examine later that the DRM-based estimator of $ G_1 (x) $ attains the same asymptotic efficiency as above.

\subsection{Quantile estimation under the parametric submodel}
\label{sec:quan_Para} 

Quantiles are very important population parameters in many applications. 
In this section, we study the asymptotic efficiency of the parametric quantile estimator under the exponential family model in \eqref {submodel}. 
We first define, for any quantile level $ p \in (0, 1) $, the $ p $th quantile of $ G_1 $ as 
\[ 
\xi_p = G_1^{-1} (p) \coloneqq \inf \{ t: G_1 (t) \geq p \}, 
\]
where the inverse function $ G_1^{-1} (\cdot) $ is also known as the quantile function. 
We assume that the distribution $ G_1 (x) $ has a density function $ g_1 (x) $ that is positive and continuous at $ x = \xi_p $. 
Then, by Theorem~\ref {invariance_MLE}, the MLE of $ \xi_p = G_1^{-1} (p; \bdeta_1^{*}) $ is given by 
\[ 
\tilde \xi_p = G_1^{-1} (p; \tilde \bdeta_1), 
\] 
where $ \tilde \bdeta_1 $ is the MLE of $ \bdeta_1 $ under the parametric model in \eqref {submodel}.

Because the MLE $ \tilde \bdeta_1 $ is asymptotically normal (see \eqref{mle-eta-normal}), the quantile MLE $ \tilde \xi_p $ is also asymptotically normal by the delta method: 
\begin{align}
\sqrt {n_1} 
\{ \tilde \xi_p - \xi_p \} 
& = 
\sqrt {n_1} 
\{ G_1^{-1} (p; \tilde \bdeta_1) - G_1^{-1} (p; \bdeta_1^{*}) \} 
\nonumber \\ 
& \overset {d} \to 
N 
\left (
\bsm {0}, 
\frac {\partial G_1^{-1} (p; \bdeta_1^{*})} {\partial \bdeta_1^{\top}}
\Var_1^{-1} [\bq_{-} (X)]
\frac {\partial G_1^{-1} (p; \bdeta_1^{*})} {\partial \bdeta_1}
\right ), 
\hspace{2mm}
\text { as } n_1 \to \infty.
\label {mle-quan-efficiency}
\end{align}
In addition, the variance matrix in \eqref {mle-quan-efficiency} has a more specific expression, as follows. 
Firstly, 
\begin{align*} 
\frac {\partial G_1^{-1} (p; \bdeta_1^{*})} {\partial \bdeta_1} 
= 
- \frac {1} {g_1 (\xi_p)} \frac {\partial G_1 (\xi_p; \bdeta_1^{*})} {\partial \bdeta_1}. 
\end{align*} 
Then, from previous derivations given in \eqref {A_derivative} and \eqref {G_derivative}, 
we have 
\begin{align*} 
\frac {\partial G_1 (\xi_p; \bdeta_1^{*})} {\partial \bdeta_1}
= \bQ (\xi_p) - G_1 (\xi_p) \bbE_1 [\bq_{-} (X)] 
= \bQ (\xi_p) - p \bbE_1 [\bq_{-} (X)]. 
\end{align*} 
Hence, the variance matrix in \eqref {mle-quan-efficiency} can be written as 
\begin{align} 
& \frac {\partial G_1^{-1} (p; \bdeta_1^{*})} {\partial \bdeta_1^{\top}}
\Var_1^{-1} [\bq_{-} (X)]
\frac {\partial G_1^{-1} (p; \bdeta_1^{*})} {\partial \bdeta_1} 
\nonumber \\ 
& = 
\left \{ \bQ (\xi_p) - p \bbE_1 [\bq_{-} (X)] \right \}^{\top}
\frac {\Var_1^{-1} [\bq_{-} (X)]} {g_1^{2} (\xi_p)}
\left \{ \bQ (\xi_p) - p \bbE_1 [\bq_{-} (X)] \right \}. 
\label {mle-quan-efficiency-expression}
\end{align} 
Our objective, to be realized later, is to investigate whether the DRM-based estimator of $ \xi_p $ has the asymptotic variance as low as the variance in \eqref{mle-quan-efficiency} and equivalently in \eqref{mle-quan-efficiency-expression}.

\section{Estimation efficiency under the DRM}
\label{sec:EL-DRM} 

Under the DRM, the base distribution $ G_0 $ is left unspecified. 
If we impose a parametric form on $ G_0 $, the DRM would reduce to a parametric model. 
In this case, we gain model simplicity but bear a higher risk of model misspecification. 
We thus leave $ G_0 $ unspecified and use the nonparametric EL of \citet{owen1988empirical} as a platform for statistical inference under the DRM. 
There has been a rich literature on the coupling of the EL and the DRM; see, for example, \citet{qin1993empirical, qin1997goodness, qin1998inferences, fokianos2001semiparametric}. 

\subsection{EL-based inference under the DRM}

We first review the EL method under the DRM. 
For convenience, let $ p_{kj} = \mathrm{d} G_0 (x_{k j}) = P (X = x_{k j}; G_0) $, the probability of observing $ x_{k j} $ under $ G_0 $ for all applicable $ k, j $. 
Applying the likelihood principle, we obtain the EL based on the multiple sample under the DRM:
\begin{align} 
L_{n} (G_{0}, \ldots, G_m) 
= 
\prod_{k, j} P (X = x_{k j}; G_{k}) 
 = 
\prod_{k=0}^m\prod_{j = 1}^{n_k} p_{kj} \exp \{ \btheta_k^{\top} \bq (x_{k j}) \}, 
\label {EL-DRM}
\end{align}
with $ \btheta_0 \coloneqq \bsm{0} $ by convention. 
We observe that $L_{n} (\cdot) = 0$ if $ G_k $ are continuous distribution functions. 
This seemingly devastating property does little harm to the usefulness of the EL. 
As we will see, in the EL we search for the distribution estimator within the space of discrete distributions that assign positive probability mass to the observed data. 
This does not eliminate much generality because every distribution can be precisely approximated by such discrete distributions when the sample size grows.  
Since $ L_{n} (\cdot)  $ in \eqref {EL-DRM} is also a function of the parameters $ \btheta $ and base distribution $ G_0 $, we may also write its logarithm as 
\begin{align} 
\ell_{n} (\btheta, G_{0}) 
= \sum_{k, j} \log p_{k j} 
+ 
\sum_{k=1}^m \sum_{j = 1}^{n_k} \btheta_k^{\top} \bq (x_{k j}). 
\label {log-EL}
\end{align}  

The EL-based inferences on the population parameters are usually carried out through a profile likelihood function. 
We first observe that the DRM assumption in \eqref{DRM} implies
\[ 
1 = \int \mathrm{d} G_0 (x), 
\hspace{5mm} 
1 = \int \mathrm{d} G_r = \int \exp \{ \btheta_r^{\top} \bq (x) \} \mathrm{d} G_0 (x), 
\hspace{3mm} 
r = 1, \ldots, m. 
\] 
Confining the common support of $ G_0, \ldots, G_m $ to the observed data $ \{ x_{k j} \}_{k, j} $ yields the EL-version constraints 
\[ 
\sum_{k, j} p_{k j} = 1,  
\hspace{5mm} 
\sum_{k, j} p_{k j} \exp \{ \btheta_r^{\top} \bq (x_{k j}) \} = 1, 
\hspace{3mm} 
r = 1, \ldots, m. 
\] 
With these preparations and following the foundational work by \citet{qin1994empirical}, we define the profile log-EL function of $ \btheta $ as the supremum of the log-EL $ \ell_{n} (\btheta, G_{0}) $ in \eqref{log-EL} over $ G_0 $ subject to the above constraints: 
\begin{align*}
\tilde \ell_{n} (\btheta) 
= \sup_{G_{0}} 
\Big \{ \ell_{n} (\btheta, G_{0}): 
\sum_{k, j} p_{k j} = 1, 
\hspace {3mm}
\sum_{k, j} p_{k j} \exp \{ \btheta_r^{\top} \bq (x_{k j}) \} = 1, 
\hspace{3mm} r = 1, \ldots, m, 
\Big \}.  
%\label {profile-log-EL}
\end{align*}
The above optimization problem has a simple solution by the Lagrange multiplier method. 
For the best relevance, we present the results when $ m = 1 $. 
We have
\begin{align*} 
\tilde \ell_{n} (\btheta) 
= 
- \sum_{k, j} \log \Big [ n + \hat \lambda_1 
\big ( \exp \{ \btheta^{\top} \bq (x_{k j}) \} - 1 \big ) \Big ] 
+ 
\sum_{j = 1}^{n_1} \btheta^{\top} \bq (x_{1 j}), 
  %\label {langeq10}  
\end{align*} 
for some $ \hat \lambda_1 $ satisfying 
$ \sum_{k, j} 1/[ n + \hat \lambda_1 ( \exp \{ \btheta^{\top} \bq (x_{k j}) \} - 1)] = 1 $. 
One may estimate $ \btheta $ by the maximum empirical likelihood estimator (MELE): 
\begin{align} 
\hat \btheta = \argmax \tilde \ell_{n} (\btheta). 
\label{MELE}
\end{align} 

At $ \btheta = \hat \btheta $, some algebra gives $ \hat \lambda_1 = n_1 $, and we naturally get another function by replacing $ \hat \lambda_1 $ with $ n_1 $ in the profile log-EL $ \tilde \ell_{n} (\btheta) $: 
\begin{align} 
\ell_{n} (\btheta) 
& = - \sum_{k, j} \log \big [ n_0 + n_1 \exp \{ \btheta^{\top} \bq (x_{k j}) \} \big ] 
+ \sum_{j = 1}^{n_1} \btheta^{\top} \bq (x_{1 j}). 
 \label {dual-log-EL}  
\end{align} 
The profile log-EL $ \tilde \ell_{n} (\btheta) $ and $ \ell_{n} (\btheta) $ have the same maximum value  and maximizer. 
Because of this, we study the asymptotic properties of the MELE $ \hat \btheta $ 
through the analytically simpler $ \ell_{n} (\btheta) $. 
By convention, we call this function a {\em dual function} of the profile log-EL. 
Following \citet{chen2013quantile}, we regard this dual function $ \ell_{n} (\btheta) $ as if it is the profile log-EL. 

Our ultimate goal is to study the asymptotic efficiency of the DRM-based inferences of $ G_1 $ that has a smaller sample under the two-sample scenario when $ n_0/n_1 \to \infty $. 
To achieve this goal, we first show that the MELE $ \hat \btheta $ under the DRM is asymptotic normal with the same low variance as the MLE $ \tilde \btheta $ in \eqref{theta-MLE} derived under the parametric submodel \eqref{submodel}.
Therefore, when the sample sizes $ n_0/n_1 \to \infty $ and the DRM holds, the inference on $ G_1 $ {\blue \bf is asymptotically as efficient} as it attains under a corresponding correct parametric model for $ G_1 $. 
We further prove that the DRM-based estimator of the cumulative distribution function $ G_1 (x) $ and the quantiles of $ G_1 $ both achieve parametric efficiency under the two-sample scenario. 
Proofs of the main results in this sectino are provided in \hyperref[sec:appendix]{Appendix}.

\subsection{Estimation of \texorpdfstring{$\boldsymbol{\theta}$}{} under the DRM}
\label{sec:para_DRM} 

In this section, we show that the MELE $ \hat \btheta $ in \eqref{MELE} under the DRM is asymptotic normal with the same low asymptotic variance as the parametric MLE $ \tilde \btheta $ in \eqref{theta-MLE} under the corresponding parametric submodel in \eqref{submodel}. 
Therefore, when the sample sizes $ n_0/n_1 \to \infty $ and the semiparametric DRM holds, the inference on $G_1$ is as efficient as we postulate a correct parametric model for $G_1$.

Our theories are established under the mild conditions as follow. 
We use $ \bbE_0 [X] $ and $ \bbE_1 [X] $ for expectations calculated when $X$ has distributions $G_0$ and $G_1$ respectively. 
We denote the true parameter value by $\btheta^*$.
\begin{enumerate}[(i)] 
\item 
\label{Condition.i}
As $n_0, n_1 \to \infty$, $n_0/n_1 \to \infty$.
\item 
\label{Condition.ii}
The matrix $ \bbE_{0} [ \bq (X) \bq^{\top} (X) ] $ is positive definite. 

\item 
\label{Condition.iii}
For $ \btheta$ in a neighbourhood of the true parameter value $ \btheta^{*} $ or of $ \bsm {0} $, we have
\[
\bbE_{0} \left [ \exp (\btheta^{\top} \bq (X)) \right ] < \infty, 
\hspace {5mm} 
\bbE_{1} \left [ \exp (\btheta^{\top} \bq (X)) \right ] < \infty. 
\]
\end{enumerate}  

Condition \eqref{Condition.ii} implies that the matrix $ \bbE_{1} [ \bq (X) \bq^{\top} (X) ] $ is also positive definite. 
With respect to both $G_0$ and $G_1$, Condition \eqref{Condition.iii} states that the moment generating function of $ \bq (X) $ exists in a neighbourhood of $ \bsm {0} $.
Hence, all finite-order moments of $ \| \bq (X) \| $ are finite. 
Similarly, $ \| \bq (X) \|^{L} \exp \{ \btheta^{* \top} \bq (X) \} $ has finite expectation for all positive $L$. 
Furthermore, when $ n $ is large enough and $ \btheta $ is in a small neighbourhood of the truth $ \btheta^{*} $, the derivatives of the log-EL $ \ell_{n} (\btheta) $ are all bounded by some polynomials of $ \| \bq (X) \| $, and therefore they are all integrable.

The main goal of this section is the efficiency of MELE $ \hat \btheta $.
The following intermediate results are helpful to comprehend the main result.

\begin{lemma} 
\label{lem:profile}
Under Conditions \eqref{Condition.i} to \eqref{Condition.iii}, we have
\begin{enumerate}

\item
$\bbE \big \{
\partial \ell_{n} (\btheta^*)/{\partial \btheta} \big \} = \bsm {0}; 
$

\item
$
n_1^{-1/2} \{ \partial \ell_{n} (\btheta^*)/ {\partial \btheta} \}
\overset {d} {\to}
N   (\bsm {0}, \Var_1   [ \bq (X)   ]  ) 
$
\hspace{1mm} as $ n_0, n_1 \to \infty $;  

\item
$ - n_1^{-1}  \{ \partial^{2} \ell_{n} (\btheta^*) /\partial \btheta \partial \btheta^{\top} \} 
\overset {p} {\to}
\bbE_{1} [ \bq (X) \bq^{\top} (X) ] 
$ 
\hspace{1mm} as $ n_0, n_1 \to \infty $. 

\end{enumerate}
\end{lemma} 

Since the profile log-EL $ \ell_n (\btheta) $ has items related to observations from both $G_0$ and $G_1$, the first expectation $\bbE$ in the above lemma is with respect to these distributions item by item.
The main result is as follows.

\begin{theorem} 
\label{thm:para_normality} 

Under Conditions \eqref{Condition.i} to \eqref{Condition.iii}, as $ n_0, n_1 \to \infty $, 
the MELE $ \hat \btheta $ defined in \eqref{MELE} is asymptotically multivariate normal: 
\[ 
\sqrt {n_1} (\hat \btheta - \btheta^{*}) 
\overset {d} {\to} 
N 
\left (\bsm {0}, 
\left \{ \mathbbm {E}_{1} \left [ \bq (X) \bq^{\top} (X) \right ] \right \}^{-1} - 
\begin{pmatrix}
1 & \bsm {0} \\ 
\bsm {0} & \bsm {0} 
\end{pmatrix}
\right ). 
\] 

\end{theorem} 

We can now answer the question: is the asymptotic variance as low as it can be under the parametric submodel in \eqref{submodel}? 
Recall that $\bq^{\top} (x) = (1, \bq_{-}^{\top} (x))$, and let $ \bI_d $ be the identity matrix of dimension $ d $. 
After some matrix algebra \citep[Theorem 8.5.11]{harville1997matrix}, we have
\begin{align}
& \Big \{ \mathbbm {E}_{1} [\bq (X) \bq^{\top} (X)] \Big \}^{-1} - 
\begin{pmatrix}
1 & \bsm {0} \\ 
\bsm {0} & \bsm {0} 
\end{pmatrix} 
\nonumber \\ 
& = 
\begin{pmatrix}
\bbE_1 [\bq_{-}^{\top} (X)] \Var_1^{-1} [\bq_{-} (X)] \bbE_1 [\bq_{-} (X)] 
& 
- \bbE_1 [\bq_{-}^{\top} (X)] \Var_1^{-1} [\bq_{-} (X)] \\ 
- \Var_1^{-1} [\bq_{-} (X)] \bbE_1 [\bq_{-} (X)]
& 
\Var_1^{-1} [\bq_{-} (X)] 
\end{pmatrix} 
\nonumber \\ 
& = 
\begin{pmatrix}
- \bbE_1 [\bq_{-}^{\top} (X)] \\ 
\bI_d
\end{pmatrix}
\Var_1^{-1} [\bq_{-} (X)]
\begin{pmatrix}
- \bbE_1 [\bq_{-} (X)], 
\, 
\bI_d
\end{pmatrix}. 
\label{thm:para_normality_mat}
\end{align} 

With the help from the identity in \eqref{thm:para_normality_mat}, we can see that the asymptotic variance in Theorem~\ref{thm:para_normality} is exactly equal to the asymptotic variance of the parametric MLE of $ \btheta $ in \eqref{mle-efficiency}. 
In other words, when $ n_0/n_1 \to \infty $, the DRM-based estimator of $ \btheta $ achieves the same asymptotic efficiency as the parametric estimator under the submodel \eqref{submodel}, namely the highest possible efficiency. 
Estimating the parameter $ \btheta $ and hence also the density ratio is an interesting problem by itself. 
We, however, focus more on the efficiency of estimating the cumulative distribution function and quantiles of $ G_1 $. 
The following sections are devoted to these tasks. 
As a note for notation, hereafter we may drop the dummy variable inside the expectation and variance operators for a better presentation: 
\[ 
\bbE_1 [\bq_{-}] = \bbE_1 [\bq_{-} (X)], 
\hspace {5mm}
\Var_1 [\bq_{-}] = \Var_1 [\bq_{-} (X)]. 
\]

\subsection{Distribution estimation under the DRM}
\label{sec:dist_DRM} 

In this section, we investigate the asymptotic efficiency of the DRM-based estimator of $ G_1 (x) $ under the two-sample scenario. 
We first define this estimator. 
With the DRM-based MELE $ \hat \btheta $ defined in \eqref{MELE}, we have the fitted values of $ p_{k j} = P (X = x_{k j}; G_0) $ that characterize $ G_0 $: 
\[ 
\hat p_{k j} = [ n_0 + n_1 \exp \{ \hat \btheta^{\top} \bq (x_{k j}) \} ]^{-1}.  
\] 
We then naturally obtain an estimator of $ G_0 (x) $ as $ \hat G_0 (x) = \sum_{k, j} \hat p_{k j} \ind (x_{k j} \leq x) $, and an estimator of $ G_1 (x) $ under the DRM: 
\begin{align}
\hat G_1 (x) = \sum_{k, j} \hat p_{k j} \exp \{ \hat \btheta^{\top} \bq (x_{k j}) \} \ind (x_{k j} \leq x), 
\label{DRM-G1}
\end{align}
where $ \ind (\cdot) $ is the indicator function. 

The following theorem states that $ \hat G_1 (x) $ is also asymptotically normal as $ n_0/n_1 \to \infty $. 

\begin{theorem} 
\label{thm:dist_normality} 

Under Conditions \eqref{Condition.i} to \eqref{Condition.iii}, 
for every $ x $ in the support of $ G_1 $, 
we have that as $ n_0, n_1 \to \infty $, 
\begin{align*} 
\sqrt {n_1} & \{ \hat G_1 (x) - G_1 (x) \} 
\overset {d} {\to} 
\\ 
& N 
\left (0, 
\{ \bQ (x) - \bbE_1 [\bq_{-}] G_1 (x) \}^{\top} 
\Var_1^{-1} [\bq_{-}] 
\{ \bQ (x) - \bbE_1 [\bq_{-}] G_1 (x) \} 
\right ), 
\end{align*}
where we define 
\[ 
\bQ (x) = \int_{-\infty}^{x} \bq_{-} (t) \mathrm {d} G_1 (t).
\] 
The variance matrix in the limiting distribution can also be written as 
\[ 
\Cov_1 [\bQ^{\top} (X), \bone (X \leq x)] 
\Var_1^{-1} [\bq_{-} (X)] 
\Cov_1 [\bQ (X), \bone (X \leq x)], 
\] 
where $ \Cov_1 $ denotes the covariance calculated under distribution $ G_1 $. 

\end{theorem} 

We observe that the asymptotic variance in the above theorem is the same as that in \eqref{mle-dist-efficiency-eqv}, which further equals the asymptotic variance of the parametric MLE $ \tilde G_1 (x) $ in \eqref{mle-dist-efficiency}. 
Therefore, same as the inference on $ \btheta $, the DRM-based estimator of the distribution function $ G_1 (x) $ is also as efficient as the parametric estimator under the submodel model in \eqref{submodel} when $ n_0/n_1 \to \infty $.

\subsection{Quantile estimation under the DRM}
\label{sec:quan_DRM} 

In this section, we derive the asymptotic variance of DRM-based quantile estimator for $ G_1 $. 
The $ p $th quantile of $ G_1 $ is denoted by $ \xi_p $. 
Estimator of $ \xi_p $ under the DRM is constructed based on the distribution estimator $ \hat G_1 (x) $ defined in \eqref{DRM-G1}:
\begin{align}
\hat \xi_p = \inf \{ t: \hat G_1 (t) \geq p \}. 
\label{DRM-quan}
\end{align}
Because of the relationship between the quantile and distribution function, we often study the asymptotic properties of quantile estimator through distribution estimator.   
With the help from Theorem~\ref {thm:dist_normality} for the asymptotic normality of the DRM-based distribution estimator, we are able to derive the limiting distribution of the DRM-based quantile estimator $ \hat \xi_p $, as shown in the following theorem. 

\begin{theorem} 
\label{thm:quan_normality}

Under Conditions \eqref{Condition.i} to \eqref{Condition.iii}, and suppose that the density function $ g_1 (\cdot) $ is continuous and positive at $ \xi_p $. 
The DRM-based quantile estimator is asymptotically normal: 
\begin{align*} 
\sqrt {n_1} (\hat \xi_p - \xi_p) 
\overset {d} \to 
N 
\left (0, 
\{ \bQ (\xi_p) - p \bbE_1 [\bq_{-}] \}^{\top} 
\frac {\Var_1^{-1} [\bq_{-}]} {g_1^2 (\xi_p)} 
\{ \bQ (\xi_p) - p \bbE_1 [\bq_{-}] \} 
\right ), 
\end{align*}
as $ n_0, n_1 \to \infty $. 

\end{theorem} 

Evidently, the asymptotic variance in Theorem~\ref {thm:quan_normality} equals the variance in \eqref{mle-quan-efficiency-expression}, and equivalently is as low as the asymptotic variance in \eqref{mle-quan-efficiency} of the parametric MLE $ \tilde \xi_p $. 
Hence, under the two-sample scenario where $ n_1/n_0 \to 0 $, we have proved that the DRM-based quantile estimator for $ G_1 $ attains parametric efficiency as if the parametric model in \eqref{submodel} is assumed.

Beyond our core contribution, we also present a useful result that reveals a linear relationship between quantile estimator and distribution estimator. 
This type of result is famously known as the Bahadur representation. 

\begin{theorem} 
\label{thm:bahadur} 

Under Conditions \eqref{Condition.i} to \eqref{Condition.iii}, and suppose that the density function $ g_1 (\cdot) $ is continuous and positive at $ \xi_p $. 
The DRM-based quantile estimator has Bahadur representation: 
\begin{align*} 
\hat \xi_p = \xi_p + \frac {G_1 (\xi_p) - \hat G_1 (\xi_p)} {g_1 (\xi_p)} 
+ O_p (n_1^{-3/4} \log^{1/2} n_1). 
\end{align*}

\end{theorem} 

The conclusion in Theorem~\ref{thm:bahadur} is stronger than that in Theorem~\ref{thm:quan_normality}.

\section{Efficiency of DRM-based quantile estimation when $ G_0 = G_1 $} 
\label{sec:DRM-identical} 

For illustration, in this section we demonstrate how the asymptotic variance of the DRM-based quantile estimator $ \hat \xi_p $ evolves as a function of the sample size ratio $ k = n_0/n_1 $ when two population distributions are actually identical. 
That is, we study the efficiency when the true model parameter $ \btheta^* = \mathbf {0} $ in the DRM \eqref{DRM}. 
However, we do not assume the knowledge of $ G_0 = G_1 $ when fitting the DRM to the data. 
Although our main focus is on the situation when $ k \to \infty $, we can learn a lot from the case when $ k $ is finite but large. 
Applying the results from \citet {zhang2000quantile} and \citet {chen2013quantile}, the following corollary explicitly quantifies the efficiency gap between the DRM-based quantile estimator $ \hat \xi_p $ and the parametric MLE $ \tilde \xi_p $ of the quantile. 

\begin{cor} 
\label{cor:ELDRM_asym_var_normal}

Assume that the DRM holds with the true $ \btheta^* = \mathbf {0} $, and $ \bbE_{0} \left [ \exp (\btheta^{\top} \bq (X)) \right ] < \infty $ for $ \btheta $ in a neighbourhood of $ \bsm{0} $.  
Further, assume that $ k = n_0/n_1 $ is fixed, finite, and positive as $ n_0, n_1 \to \infty $, and that the density function $ g_1 (x) $ is continuous and positive at $ x = \xi_p $. 
Then, the centralized DRM-based quantile estimator $ \sqrt{n_1} (\hat \xi_p - \xi_p) $ has a limiting normal distribution with variance 
\begin{align} 
\sigma_{\hat \xi_{p}}^2
= \frac {1} {k + 1} \frac {p (1 - p)} {g_{1}^{2} (\xi_{p})}  
+ \frac {k} {k + 1} \sigma_{\tilde \xi_{p}}^2, 
\label {ELDRM_asym_var_normal}
\end{align} 
where $ \sigma_{\tilde \xi_{p}}^2 $ is the variance of the limiting normal distribution of the MLE $ \tilde \xi_p $ that has a matrix expression given in \eqref {mle-quan-efficiency-expression}. 

\end{cor}

The asymptotic variance in \eqref{ELDRM_asym_var_normal} depends on $ k = n_0/n_1 $, providing an insight on how the efficiency of the DRM quantile estimator evolves as $ n_0/n_1 \to \infty $.  
We now take a closer look at the first part in the right hand side of \eqref{ELDRM_asym_var_normal}. 
Let 
$
F_{n, 1} (x) = n_1^{-1} \sum_{j=1}^{n_1} \ind (X_{1 j} \leq x) 
$
denote the empirical distribution of the sample from $ G_1 $. 
The empirical quantile with level $ p $ for $ G_1 $ is  
$
\bar \xi_{p} = \inf \{ x: F_{n, 1} (x) \geq p \}. 
$
It is well known that if $ g_1 (\xi_p) > 0 $, $ \bar \xi_{p} $ is asymptotically normal \citep{serfling2000approximation}:  
\begin{align} 
\sqrt {n_1} (\bar \xi_{p} - \xi_{p})
\overset {d} \to N 
\left (0, \frac {p (1-p)} {g_1^2 (\xi_{p})} \right ), 
\hspace{3mm} 
\text { as } n_1 \to \infty.      
\label{sample_quantile_asym_normal}
\end{align} 

Therefore, the asymptotic variance of the DRM-based quantile estimator given in \eqref {ELDRM_asym_var_normal} is a weighted average of the asymptotic variance of the empirical quantile and the asymptotic variance of the MLE. 
Interestingly, as $ k = n_0/n_1 $ increases, the efficiency of the DRM-based quantile estimator $ \hat \xi_{p} $ approaches the efficiency of the MLE $ \tilde \xi_p $. 
Although this intriguing observation is developed under a special situation where $ G_0 = G_1 $ and $ k $ is assumed to be fixed and positive, it enlightens us on the efficiency of the DRM-based quantile estimator in addition to our theoretical result in Theorem \ref{thm:quan_normality} that is established under general conditions and evolving $ k $.

\section{Simulation studies} 
\label{sec:simulations}

In this section, we report some simulation results on the efficiency of the DRM-based estimators of quantiles. 
We are particularly interested in their efficiency in estimating lower or higher quantiles, compared to the empirical quantiles and the parametric MLEs. 
Because the probability density function often has small value at $\xi_p$ when $p$ is close to zero or one, the corresponding empirical quantile has very large variance according to \eqref{sample_quantile_asym_normal}. 
In contrast, the parametric quantile estimators are largely free from this deficiency but suffer from serious bias when the model is misspecified.
Based on our theoretical results, DRM-based quantile estimator achieves the efficiency of the parametric estimator with low model misspecification risk in the presence of an extra sample with a much larger size. 
In the following sections, we use 1000 repetitions to obtain the simulated biases and variances of the three quantile estimators: the DRM-based estimator, the parametric MLE, and the empirical quantile.

\subsection{Data generated from normal distributions} 
\label{ch:DRM-eff-sec:DRM-Normal} 

We first examine the performance of the DRM-based quantile estimator when data are from normal distributions. 
The family of normal distributions has many nice statistical properties. 
For example, the MLEs of the model parameters and quantiles under the normal family have closed forms. 
Such properties make it easier to compare the efficiency between the DRM-based estimator and the parametric MLE. 
Also, the normal distributions collectively satisfy the DRM with basis function $ \bq (x) = (1, x, x^2)^{\top} $.
We assume the knowledge of this basis function but not the parametric form in the DRM approach in simulations. 

We generate the first sample from $ N (\mu_{0}, \sigma_{0}^{2}) $ and the second sample from $ N (\mu_{1}, \sigma_{1}^{2}) $. 
We observe the performance of the DRM-based quantile estimators for various choices of the means $ \mu_0, \mu_1 $, standard deviations $ \sigma_0, \sigma_1 $, and sample sizes $ n_0, n_1 $. 
Under normal model, the parametric MLE quantile estimator of $\xi_p$ is given by 
\begin{align*} 
\tilde \xi_{p} =  \tilde \mu_1 +  \Phi^{-1} (p) \tilde \sigma_1, 
\end{align*} 
where $\Phi^{-1} (p)$ is the $p$th quantile of the standard normal distribution, and $\tilde \mu_1$ and $\tilde \sigma_1$ are the sample mean and standard deviation based on $\{x_{1 j}: j=1, \ldots, n_1\}$. 
It can be shown that the asymptotic variance of the MLE $ \tilde \xi_{p} $ is given by 
\begin{align*} 
n_1 \Var (\tilde \xi_{p}) 
= \sigma_1^2 \left \{ 1 + \frac { [\Phi^{-1} (p)]^2 } { 2 } \right \}.    
%\label {var_mle_quantile4}   
\end{align*} 

We first generate both samples from standard normal distribution, namely, $ \mu_0 = \mu_1 = 0 $ and $ \sigma_0 = \sigma_1 = 1 $. 
Table~\ref{eff_quan1_1} contains the simulated biases and variances of the three quantile estimators after being properly scaled based on sample size. 
We consider 4 quantile levels $ p \in \{ 0.01, 0.05, 0.10, 0.50 \} $ and 4 sample size combinations $ n_1 \in \{ 100, 1000 \} $ and $ n_0 \in \{ 10 n_1, 100 n_1 \} $. 
Due to symmetry of the normal distribution, we do not include the higher level quantiles.
The biases and variances in the table are inflated by a factor $ \sqrt {n_1} $ and $ n_1 $ respectively.
We observe that when $ n_0/n_1 = 100 $, the asymptotic variances of the DRM-based quantile estimators approximately equal the weighted averages of the variances of the parametric MLEs (with weight $ k/(k+1) $) and empirical quantiles (with weight $ 1/(k+1) $). 
This is consistent with our theoretical finding in Corollary \ref{cor:ELDRM_asym_var_normal}. 
Further, when $n_0/n_1$ increases from $10$ to $100$, the variances of the DRM-based quantile estimators $ \hat \xi_{p} $ rapidly approaches the variances of the parametric MLEs $ \tilde \xi_{p} $. 
The improvement in efficiency due to increased $n_0/n_1$ is especially significant for quantiles at levels $p$ close to zero. 

We next experiment with data from normal distributions with $ \mu_0 = 1, \mu_1 = 2 $ and $ \sigma_0^2 = 1.5, \sigma_1^2 = 2 $. 
The performance of the three quantile estimators is summarized in Table~\ref {eff_quan1_2}. 
We note that our previous comments on the efficiency of the DRM-based estimators when $ n_0/n_1 $ increases are also applicable here, and therefore we do not offer more interpretations. 
We anticipate other combinations of normal distribution, if not far different, will still lead to similar results. 

\begin{table}[h!] 
\begingroup
\setlength{\tabcolsep}{6pt} % Default value: 6pt
\renewcommand{\arraystretch}{1} % Default value: 1
\caption{Simulated bias ($ \times \sqrt {n_1} $) and variance ($ \times n_1 $) of quantile estimators. Both samples are from standard normal.}
\label{eff_quan1_1}
\begin{center}
\begin{adjustbox}{max width=\textwidth, totalheight = \textheight*3/5}
\begin{tabular}{lrrrrrrrr}
\hline
\multicolumn{1}{l}{Level $ p $}         
& \multicolumn{2}{c}{DRM-based}       
& \multicolumn{2}{c}{Para MLE}  
& \multicolumn{2}{c}{Empirical}  
\\ 
\cmidrule(lr){2-3} \cmidrule(lr){4-5} \cmidrule(lr){6-7}
& Bias & Var & Bias & Var & Bias & Var \\ 
\hline
& \multicolumn{6}{c}{$n_{1} = 100, ~~~ n_0 = 10 n_1, $} \\ 
\cline{2-7}
0.01      &  
$ 0.11 $ & $ 5.19 $ & $ 0.25 $ & $ 3.73 $ & $ 1.97 $ & $ 10.44 $
\\    
0.05      &  
$ 0.12 $ & $ 2.61 $ & $ 0.18 $ & $ 2.36 $ & $ 0.60 $ & $ 4.20 $
\\   
0.10      &  
$ 0.11 $ & $ 1.97 $ & $ 0.15 $ & $ 1.82 $ & $ 0.33 $ & $ 2.77 $
\\   
0.50      &  
$ 0.01 $ & $ 1.04 $ & $ 0.03 $ & $ 0.99 $ & $ 0.01 $ & $ 1.44 $
\\   
%0.90      &  
%$ -0.11 $ & $ 1.89 $ & $ -0.09 $ & $ 1.81 $ & $ -0.27 $ & $ 2.85 $
%\\   
%0.95      &  
%$ -0.14 $ & $ 2.55 $ & $ -0.13 $ & $ 2.34 $ & $ -0.43 $ & $ 4.39 $
%\\    
& \multicolumn{6}{c}{$n_{1} = 100, ~~~ n_0 = 100 n_1 $} \\ 
\cline{2-7}
0.01      &  
$ 0.17 $ & $ 3.73 $ & $ 0.18 $ & $ 3.57 $ & $ 1.90 $ & $ 8.91 $
\\    
0.05      &  
$ 0.12 $ & $ 2.29 $ & $ 0.13 $ & $ 2.27 $ & $ 0.42 $ & $ 4.28 $
\\   
0.10      &  
$ 0.10 $ & $ 1.77 $ & $ 0.10 $ & $ 1.75 $ & $ 0.21 $ & $ 2.63 $
\\   
0.50      &  
$ -0.01 $ & $ 0.96 $ & $ -0.01 $ & $ 0.96 $ & $ 0.02 $ & $ 1.48 $
\\   
%0.90      &  
%$ -0.11 $ & $ 1.73 $ & $ -0.11 $ & $ 1.72 $ & $ -0.24 $ & $ 2.71 $
%\\   
%0.95      &  
%$ -0.14 $ & $ 2.28 $ & $ -0.14 $ & $ 2.23 $ & $ -0.46 $ & $ 4.20 $
%\\    
& \multicolumn{6}{c}{$n_{1} = 1000, ~~~ n_0 = 10 n_1 $} \\ 
\cline{2-7}
0.01      &  
$ -0.02 $ & $ 4.91 $ & $ 0.03 $ & $ 3.81 $ & $ 0.51 $ & $ 13.53 $
\\    
0.05      &  
$ -0.01 $ & $ 2.61 $ & $ 0.02 $ & $ 2.42 $ & $ 0.09 $ & $ 4.53 $
\\   
0.10      &  
$ 0.00 $ & $ 1.98 $ & $ 0.02 $ & $ 1.87 $ & $ 0.03 $ & $ 3.29 $
\\   
0.50      &  
$ 0.00 $ & $ 1.10 $ & $ 0.01 $ & $ 1.03 $ & $ 0.05 $ & $ 1.58 $
\\   
%0.90      &  
%$ -0.01 $ & $ 1.98 $ & $ 0.00 $ & $ 1.88 $ & $ -0.05 $ & $ 2.89 $
%\\   
%0.95      &  
%$ 0.00 $ & $ 2.56 $ & $ 0.00 $ & $ 2.43 $ & $ -0.08 $ & $ 4.42 $
%\\   
& \multicolumn{6}{c}{$n_{1} = 1000, ~~~ n_0 = 100 n_1 $} \\ 
\cline{2-7}
0.01      &  
$ -0.06 $ & $ 3.94 $ & $ -0.05 $ & $ 3.83 $ & $ 0.58 $ & $ 13.66 $
\\    
0.05      &  
$ -0.05 $ & $ 2.46 $ & $ -0.05 $ & $ 2.41 $ & $ 0.11 $ & $ 4.45 $
\\   
0.10      &  
$ -0.05 $ & $ 1.86 $ & $ -0.05 $ & $ 1.85 $ & $ 0.01 $ & $ 2.88 $
\\   
0.50      &  
$ -0.05 $ & $ 0.97 $ & $ -0.04 $ & $ 0.96 $ & $ -0.04 $ & $ 1.52 $
\\   
%0.90      &  
%$ -0.04 $ & $ 1.78 $ & $ -0.04 $ & $ 1.78 $ & $ -0.08 $ & $ 2.76 $
%\\   
%0.95      &  
%$ -0.04 $ & $ 2.31 $ & $ -0.04 $ & $ 2.32 $ & $ -0.15 $ & $ 4.26 $
%\\   
\hline
\end{tabular}
\end{adjustbox}
\end{center}
\endgroup
\end{table}

\begin{table}[h!] 
\begingroup
\setlength{\tabcolsep}{6pt} % Default value: 6pt
\renewcommand{\arraystretch}{1} % Default value: 1
\caption{Simulated bias ($ \times \sqrt {n_1} $) and variance ($ \times n_1 $) of quantile estimators. Both samples are from normal.}
\label{eff_quan1_2}
\begin{center}
\begin{adjustbox}{max width=\textwidth, totalheight = \textheight*3/5}
\begin{tabular}{lrrrrrrrr}
\hline
\multicolumn{1}{l}{Level $ p $}         
& \multicolumn{2}{c}{DRM-based}       
& \multicolumn{2}{c}{Para MLE}  
& \multicolumn{2}{c}{Empirical}  
\\ 
\cmidrule(lr){2-3} \cmidrule(lr){4-5} \cmidrule(lr){6-7}
& Bias & Var & Bias & Var & Bias & Var \\ 
\hline
& \multicolumn{6}{c}{$n_{1} = 100, ~~~ n_0 = 10 n_1 $} \\ 
\cline{2-7}
0.01      &  
$ 0.12 $ & $ 9.69 $ & $ 0.35 $ & $ 7.45 $ & $ 2.78 $ & $ 20.88 $
\\    
0.05      &  
$ 0.16 $ & $ 5.27 $ & $ 0.26 $ & $ 4.72 $ & $ 0.85 $ & $ 8.41 $
\\   
0.10      &  
$ 0.18 $ & $ 3.87 $ & $ 0.21 $ & $ 3.65 $ & $ 0.47 $ & $ 5.53 $
\\   
0.50      &  
$ 0.06 $ & $ 2.16 $ & $ 0.04 $ & $ 1.98 $ & $ 0.02 $ & $ 2.88 $
\\   
%0.90      &  
%$ -0.26 $ & $ 4.26 $ & $ -0.13 $ & $ 3.62 $ & $ -0.38 $ & $ 5.69 $
%\\   
%0.95      &  
%$ -0.47 $ & $ 6.42 $ & $ -0.18 $ & $ 4.68 $ & $ -0.61 $ & $ 8.77 $
%\\   
& \multicolumn{6}{c}{$n_{1} = 100, ~~~ n_0 = 100 n_1 $} \\ 
\cline{2-7}
0.01      &  
$ 0.16 $ & $ 7.60 $ & $ 0.26 $ & $ 7.13 $ & $ 2.68 $ & $ 17.83 $
\\    
0.05      &  
$ 0.13 $ & $ 4.69 $ & $ 0.18 $ & $ 4.53 $ & $ 0.59 $ & $ 8.56 $
\\   
0.10      &  
$ 0.11 $ & $ 3.57 $ & $ 0.14 $ & $ 3.51 $ & $ 0.30 $ & $ 5.27 $
\\   
0.50      &  
$ 0.02 $ & $ 1.95 $ & $ -0.01 $ & $ 1.91 $ & $ 0.02 $ & $ 2.96 $
\\   
%0.90      &  
%$ -0.15 $ & $ 3.59 $ & $ -0.15 $ & $ 3.45 $ & $ -0.33 $ & $ 5.42 $
%\\   
%0.95      &  
%$ -0.25 $ & $ 4.71 $ & $ -0.20 $ & $ 4.46 $ & $ -0.65 $ & $ 8.40 $
%\\   
& \multicolumn{6}{c}{$n_{1} = 1000, ~~~ n_0 = 10 n_1 $} \\ 
\cline{2-7}
0.01      &  
$ -0.11 $ & $ 9.84 $ & $ 0.04 $ & $ 7.63 $ & $ 0.73 $ & $ 27.06 $
\\    
0.05      &  
$ -0.01 $ & $ 5.54 $ & $ 0.03 $ & $ 4.84 $ & $ 0.13 $ & $ 9.06 $
\\   
0.10      &  
$ 0.01 $ & $ 4.11 $ & $ 0.02 $ & $ 3.75 $ & $ 0.04 $ & $ 6.58 $
\\   
0.50      &  
$ 0.03 $ & $ 2.24 $ & $ 0.01 $ & $ 2.07 $ & $ 0.07 $ & $ 3.16 $
\\   
%0.90      &  
%$ -0.02 $ & $ 4.51 $ & $ 0.00 $ & $ 3.76 $ & $ -0.08 $ & $ 5.77 $
%\\   
%0.95      &  
%$ -0.08 $ & $ 6.61 $ & $ 0.00 $ & $ 4.86 $ & $ -0.11 $ & $ 8.84 $
%\\   
& \multicolumn{6}{c}{$n_{1} = 1000, ~~~ n_0 = 100 n_1 $} \\ 
\cline{2-7}
0.01      &  
$ -0.10 $ & $ 8.27 $ & $ -0.07 $ & $ 7.65 $ & $ 0.82 $ & $ 27.32 $
\\    
0.05      &  
$ -0.08 $ & $ 5.03 $ & $ -0.07 $ & $ 4.81 $ & $ 0.16 $ & $ 8.90 $
\\   
0.10      &  
$ -0.07 $ & $ 3.80 $ & $ -0.06 $ & $ 3.69 $ & $ 0.02 $ & $ 5.76 $
\\   
0.50      &  
$ -0.05 $ & $ 1.94 $ & $ -0.06 $ & $ 1.92 $ & $ -0.05 $ & $ 3.04 $
\\   
%0.90      &  
%$ -0.04 $ & $ 3.70 $ & $ -0.06 $ & $ 3.55 $ & $ -0.11 $ & $ 5.52 $
%\\   
%0.95      &  
%$ -0.08 $ & $ 5.08 $ & $ -0.06 $ & $ 4.63 $ & $ -0.21 $ & $ 8.52 $
%\\   
\hline
\end{tabular}
\end{adjustbox}
\end{center}
\endgroup
\end{table}

\subsection{Data generated from exponential distributions} 

We then examine the performance of the three quantile estimators based on data from exponential distributions. 
The exponential distributions collectively fit into the DRM with the basis function $ \bq (x) = (1, x)^{\top} $.
We assume the knowledge of this $ \bq (x) $ but not the parametric form when fitting the DRM to the data in simulations. 

The quantile of the exponential distribution with mean $ \mu $ has a simple form:  $\xi_{p} = -\mu \log(1-p)$.
In the two-sample setting of this article, the parametric MLE of the second population mean is the corresponding sample mean: $\tilde \mu= n_1^{-1} \sum_{j = 1}^{n_1} x_{1 j}$.
This leads to the parametric MLE quantile estimator: 
\begin{align*} 
\tilde \xi_{p} = -\tilde \mu \log(1-p), 
\end{align*} 
with variance 
\begin{align*} 
n_1 \Var (\tilde \xi_{p}) 
= \mu^2 \log^2(1-p).    
\end{align*}

We first generate both samples from exponential distributions with $ \mu= 1 $.
We include the three quantile estimators at levels $p \in \{ 0.5, 0.9, 0.95, 0.99 \}$. 
The exponential distribution has low density values at higher level quantiles. 
Therefore, the efficiency gain at higher level quantiles are more meaningful for our investigation. 
We use the same sample size combinations as in the previous section, and the results are given in Table~\ref{eff_quan2_1}. 
We again observe the phenomenon illustrated in Corollary \ref{cor:ELDRM_asym_var_normal}: 
the asymptotic variances of the DRM-based quantile estimators are close to the weighted averages of the variances of the parametric MLEs (with weight $ k/(k+1) $) and empirical quantiles (with weight $ 1/(k+1) $). 
Same as in the normal data example, when $n_0/n_1$ increases from $10$ to $100$, the efficiency of the DRM-based quantile estimator quickly approaches the efficiency of the parametric MLE. 
The improvement is particularly obvious for quantiles at levels $p$ close to one. 

We next simulate data from exponential distributions with $ \mu_0 = 1/0.3$ and $\mu_1 = 2 $.
The performance of the quantile estimators is summarized in Table~\ref{eff_quan2_2}. 
The same conclusions regarding the DRM efficiency with growing $ n_0/n_1 $ can be drawn in this situation.
We expect similar findings for other combinations of exponential distributions.

\begin{table}[h!] 
\begingroup
\setlength{\tabcolsep}{6pt} % Default value: 6pt
\renewcommand{\arraystretch}{1} % Default value: 1
\caption{Simulated bias ($ \times \sqrt {n_1} $) and variance ($ \times n_1 $) 
of quantile estimators. Both samples are from identical exponential.}
\label{eff_quan2_1}
\begin{center}
\begin{adjustbox}{max width=\textwidth, totalheight = \textheight*3/5}
\begin{tabular}{lrrrrrrrr}
\hline
\multicolumn{1}{l}{Level $ p $}         
& \multicolumn{2}{c}{DRM-based}       
& \multicolumn{2}{c}{Para MLE}  
& \multicolumn{2}{c}{Empirical}  
\\ 
\cmidrule(lr){2-3} \cmidrule(lr){4-5} \cmidrule(lr){6-7}
& Bias & Var & Bias & Var & Bias & Var \\ 
\hline
& \multicolumn{6}{c}{$n_{1} = 100, ~~~ n_0 = 10 n_1 $} \\ 
\cline{2-7}
%0.01      &  
%$ 0.00 $ & $ 0.00 $ & $ 0.00 $ & $ 0.00 $ & $ 0.11 $ & $ 0.02 $
%\\    
%0.05      &  
%$ 0.00 $ & $ 0.01 $ & $ 0.00 $ & $ 0.00 $ & $ 0.10 $ & $ 0.06 $
%\\   
%0.10      &  
%$ 0.00 $ & $ 0.02 $ & $ 0.00 $ & $ 0.01 $ & $ 0.10 $ & $ 0.12 $
%\\   
0.50      &  
$ 0.01 $ & $ 0.55 $ & $ 0.01 $ & $ 0.50 $ & $ 0.05 $ & $ 0.98 $
\\   
0.90      &  
$ 0.02 $ & $ 6.06 $ & $ 0.03 $ & $ 5.57 $ & $ -0.33 $ & $ 9.09 $
\\   
0.95      &  
$ -0.00 $ & $ 10.49 $ & $ 0.04 $ & $ 9.43 $ & $ -0.75 $ & $ 18.68 $
\\ 
0.99      &  
$ -0.58 $ & $ 25.40 $ & $ 0.07 $ & $ 22.27 $ & $ -4.17 $ & $ 64.04 $
\\
& \multicolumn{6}{c}{$n_{1} = 100, ~~~ n_0 = 100 n_1 $} \\ 
\cline{2-7}
%0.01      &  
%$ 0.00 $ & $ 0.00 $ & $ 0.00 $ & $ 0.00 $ & $ 0.10 $ & $ 0.02 $
%\\    
%0.05      &  
%$ 0.00 $ & $ 0.00 $ & $ 0.00 $ & $ 0.00 $ & $ 0.09 $ & $ 0.07 $
%\\   
%0.10      &  
%$ 0.00 $ & $ 0.01 $ & $ 0.00 $ & $ 0.01 $ & $ 0.09 $ & $ 0.12 $
%\\   
0.50      &  
$ -0.02 $ & $ 0.50 $ & $ -0.02 $ & $ 0.50 $ & $ 0.06 $ & $ 1.01 $
\\   
0.90      &  
$ -0.06 $ & $ 5.53 $ & $ -0.06 $ & $ 5.48 $ & $ -0.45 $ & $ 8.66 $
\\   
0.95      &  
$ -0.08 $ & $ 9.53 $ & $ -0.08 $ & $ 9.28 $ & $ -0.79 $ & $ 17.70 $
\\ 
0.99      &  
$ -0.19 $ & $ 23.08 $ & $ -0.12 $ & $ 21.92 $ & $ -3.76 $ & $ 74.20 $
\\ 
& \multicolumn{6}{c}{$n_{1} = 1000, ~~~ n_0 = 10 n_1$} \\ 
\cline{2-7}
%0.01      &  
%$ 0.00 $ & $ 0.00 $ & $ 0.00 $ & $ 0.00 $ & $ 0.03 $ & $ 0.01 $
%\\    
%0.05      &  
%$ 0.00 $ & $ 0.01 $ & $ 0.00 $ & $ 0.00 $ & $ 0.03 $ & $ 0.05 $
%\\   
%0.10      &  
%$ 0.01 $ & $ 0.02 $ & $ 0.00 $ & $ 0.01 $ & $ 0.03 $ & $ 0.11 $
%\\   
0.50      &  
$ 0.03 $ & $ 0.54 $ & $ 0.03 $ & $ 0.49 $ & $ 0.05 $ & $ 0.98 $
\\   
0.90      &  
$ 0.11 $ & $ 5.54 $ & $ 0.11 $ & $ 5.42 $ & $ -0.01 $ & $ 9.00 $
\\   
0.95      &  
$ 0.12 $ & $ 10.11 $ & $ 0.14 $ & $ 9.18 $ & $ -0.21 $ & $ 19.57 $
\\   
0.99      &  
$ 0.06 $ & $ 29.28 $ & $ 0.22 $ & $ 21.68 $ & $ -1.54 $ & $ 88.77 $
\\ 
& \multicolumn{6}{c}{$n_{1} = 1000, ~~~ n_0 = 100 n_1$} \\ 
\cline{2-7}
%0.01      &  
%$ 0.00 $ & $ 0.00 $ & $ 0.00 $ & $ 0.00 $ & $ 0.03 $ & $ 0.01 $
%\\    
%0.05      &  
%$ 0.00 $ & $ 0.00 $ & $ 0.00 $ & $ 0.00 $ & $ 0.03 $ & $ 0.05 $
%\\   
%0.10      &  
%$ 0.00 $ & $ 0.01 $ & $ 0.00 $ & $ 0.01 $ & $ 0.03 $ & $ 0.11 $
%\\   
0.50      &  
$ -0.01 $ & $ 0.46 $ & $ -0.01 $ & $ 0.46 $ & $ -0.02 $ & $ 0.99 $
\\   
0.90      &  
$ -0.05 $ & $ 5.09 $ & $ -0.05 $ & $ 5.08 $ & $ -0.19 $ & $ 8.22 $
\\   
0.95      &  
$ -0.07 $ & $ 8.63 $ & $ -0.06 $ & $ 8.60 $ & $ -0.13 $ & $ 19.14 $
\\   
0.99      &  
$ -0.08 $ & $ 21.81 $ & $ -0.10 $ & $ 20.31 $ & $ -1.29 $ & $ 88.06 $
\\ 
\hline
\end{tabular}
\end{adjustbox}
\end{center}
\endgroup
\end{table}

\begin{table}[h!] 
\begingroup
\setlength{\tabcolsep}{6pt} % Default value: 6pt
\renewcommand{\arraystretch}{1} % Default value: 1
\caption{Simulated bias ($ \times \sqrt {n_1} $) and variance ($ \times n_1 $) 
of quantile estimators. Both samples are from exponential.}
\label{eff_quan2_2}
\begin{center}
\begin{adjustbox}{max width=\textwidth, totalheight = \textheight*3/5}
\begin{tabular}{lrrrrrrrr}
\hline
\multicolumn{1}{l}{Level $ p $}         
& \multicolumn{2}{c}{DRM-based}       
& \multicolumn{2}{c}{Para MLE}  
& \multicolumn{2}{c}{Empirical}  
\\ 
\cmidrule(lr){2-3} \cmidrule(lr){4-5} \cmidrule(lr){6-7}
& Bias & Var & Bias & Var & Bias & Var \\ 
\hline
& \multicolumn{6}{c}{$n_{1} = 100, ~~~ n_0 = 10 n_1$} \\ 
\cline{2-7}
%0.01      &  
%$ 0.00 $ & $ 0.01 $ & $ 0.00 $ & $ 0.00 $ & $ 0.22 $ & $ 0.09 $
%\\    
%0.05      &  
%$ 0.00 $ & $ 0.04 $ & $ 0.00 $ & $ 0.01 $ & $ 0.20 $ & $ 0.24 $
%\\   
%0.10      &  
%$ 0.00 $ & $ 0.10 $ & $ 0.00 $ & $ 0.05 $ & $ 0.19 $ & $ 0.47 $
%\\   
0.50      &  
$ -0.01 $ & $ 2.13 $ & $ 0.02 $ & $ 2.02 $ & $ 0.11 $ & $ 3.93 $
\\   
0.90      &  
$ -0.02 $ & $ 24.17 $ & $ 0.07 $ & $ 22.27 $ & $ -0.66 $ & $ 36.34 $
\\   
0.95      &  
$ -0.04 $ & $ 41.19 $ & $ 0.08 $ & $ 37.70 $ & $ -1.50 $ & $ 74.72 $
\\   
0.99      &  
$ -0.13 $ & $ 100.11 $ & $ 0.13 $ & $ 89.09 $ & $ -8.35 $ & $ 256.15 $
\\ 
& \multicolumn{6}{c}{$n_{1} = 100, ~~~ n_0 = 100 n_1 $} \\ 
\cline{2-7}
%0.01      &  
%$ 0.00 $ & $ 0.00 $ & $ 0.00 $ & $ 0.00 $ & $ 0.20 $ & $ 0.08 $
%\\    
%0.05      &  
%$ 0.00 $ & $ 0.01 $ & $ 0.00 $ & $ 0.01 $ & $ 0.19 $ & $ 0.26 $
%\\   
%0.10      &  
%$ 0.00 $ & $ 0.05 $ & $ -0.01 $ & $ 0.05 $ & $ 0.17 $ & $ 0.49 $
%\\   
0.50      &  
$ -0.04 $ & $ 2.01 $ & $ -0.04 $ & $ 1.99 $ & $ 0.13 $ & $ 4.04 $
\\   
0.90      &  
$ -0.14 $ & $ 21.85 $ & $ -0.12 $ & $ 21.92 $ & $ -0.91 $ & $ 34.64 $
\\   
0.95      &  
$ -0.16 $ & $ 37.42 $ & $ -0.16 $ & $ 37.10 $ & $ -1.57 $ & $ 70.78 $
\\   
0.99      &  
$ -0.27 $ & $ 89.22 $ & $ -0.24 $ & $ 87.68 $ & $ -7.52 $ & $ 296.81 $
\\ 
& \multicolumn{6}{c}{$n_{1} = 1000, ~~~ n_0 = 10 n_1$} \\ 
\cline{2-7}
%0.01      &  
%$ 0.00 $ & $ 0.01 $ & $ 0.00 $ & $ 0.00 $ & $ 0.06 $ & $ 0.04 $
%\\    
%0.05      &  
%$ 0.00 $ & $ 0.04 $ & $ 0.00 $ & $ 0.01 $ & $ 0.05 $ & $ 0.22 $
%\\   
%0.10      &  
%$ 0.01 $ & $ 0.11 $ & $ 0.01 $ & $ 0.05 $ & $ 0.06 $ & $ 0.43 $
%\\   
0.50      &  
$ 0.07 $ & $ 2.16 $ & $ 0.07 $ & $ 1.96 $ & $ 0.11 $ & $ 3.91 $
\\   
0.90      &  
$ 0.17 $ & $ 22.64 $ & $ 0.22 $ & $ 21.68 $ & $ -0.02 $ & $ 36.01 $
\\   
0.95      &  
$ 0.15 $ & $ 38.85 $ & $ 0.28 $ & $ 36.70 $ & $ -0.42 $ & $ 78.29 $
\\   
0.99      &  
$ 0.27 $ & $ 94.78 $ & $ 0.43 $ & $ 86.73 $ & $ -3.08 $ & $ 355.08 $
\\ 
& \multicolumn{6}{c}{$n_{1} = 1000, ~~~ n_0 = 100 n_1 $} \\ 
\cline{2-7}
%0.01      &  
%$ 0.00 $ & $ 0.00 $ & $ 0.00 $ & $ 0.00 $ & $ 0.06 $ & $ 0.05 $
%\\    
%0.05      &  
%$ 0.00 $ & $ 0.01 $ & $ 0.00 $ & $ 0.01 $ & $ 0.06 $ & $ 0.22 $
%\\   
%0.10      &  
%$ -0.01 $ & $ 0.05 $ & $ 0.00 $ & $ 0.04 $ & $ 0.05 $ & $ 0.43 $
%\\   
0.50      &  
$ -0.02 $ & $ 1.86 $ & $ -0.03 $ & $ 1.84 $ & $ -0.05 $ & $ 3.97 $
\\   
0.90      &  
$ -0.11 $ & $ 20.34 $ & $ -0.10 $ & $ 20.31 $ & $ -0.38 $ & $ 32.88 $
\\   
0.95      &  
$ -0.15 $ & $ 34.68 $ & $ -0.13 $ & $ 34.38 $ & $ -0.26 $ & $ 76.55 $
\\   
0.99      &  
$ -0.22 $ & $ 81.58 $ & $ -0.20 $ & $ 81.25 $ & $ -2.58 $ & $ 352.23 $
\\ 
\hline
\end{tabular}
\end{adjustbox}
\end{center}
\endgroup
\end{table}

\section{Real-data analysis} 
\label{sec:realdata} 

\begin{figure}[h!] 
\centering 
\includegraphics[height = 0.5\textwidth, width = \textwidth]{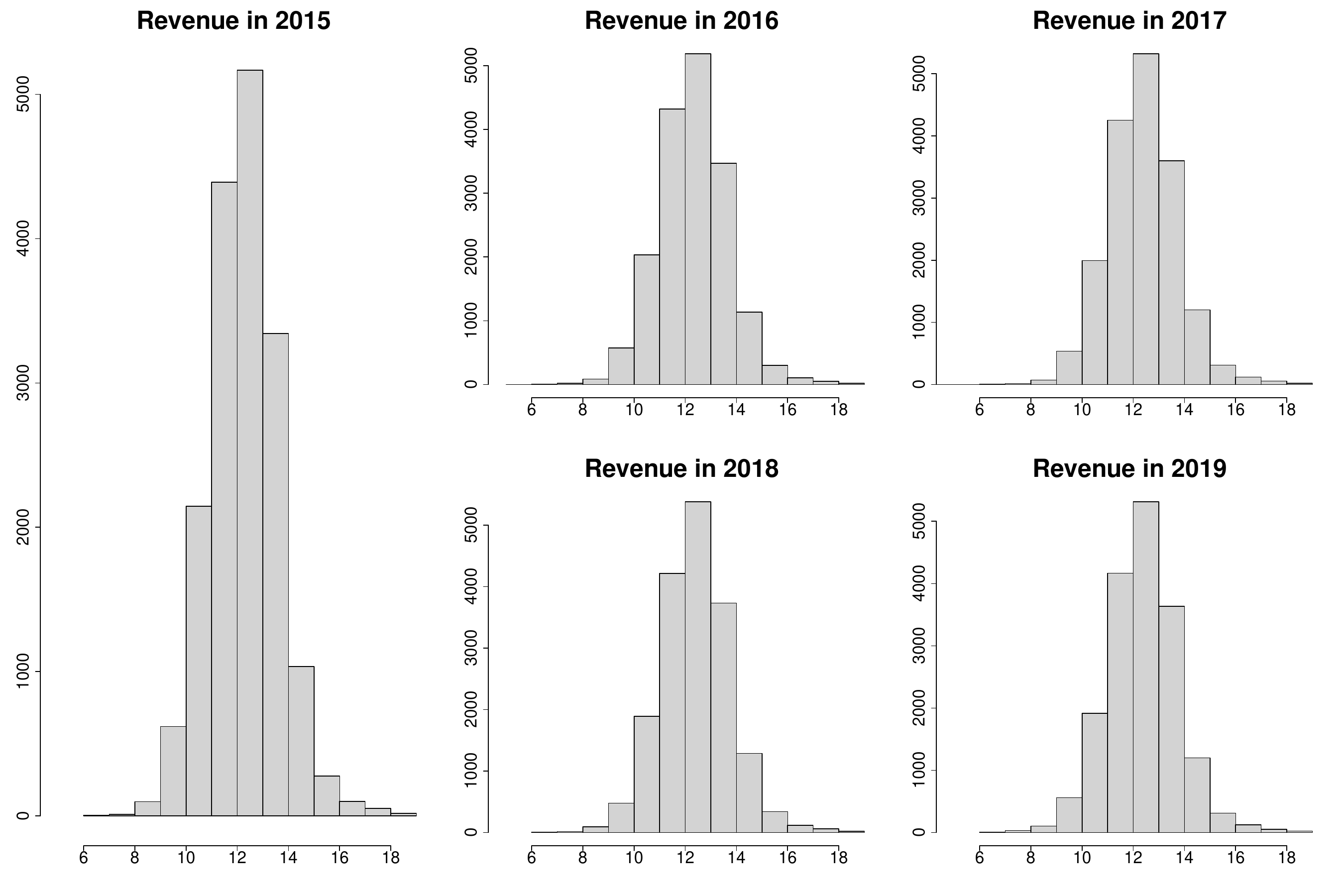} 
\caption{Histograms of log-transformed total revenues for 2015--2019.} 
\label{realdata_hist}
\end{figure}

\begin{figure}[h!]
\centering 
\includegraphics[height = 0.5\textwidth, width = 0.8\textwidth]{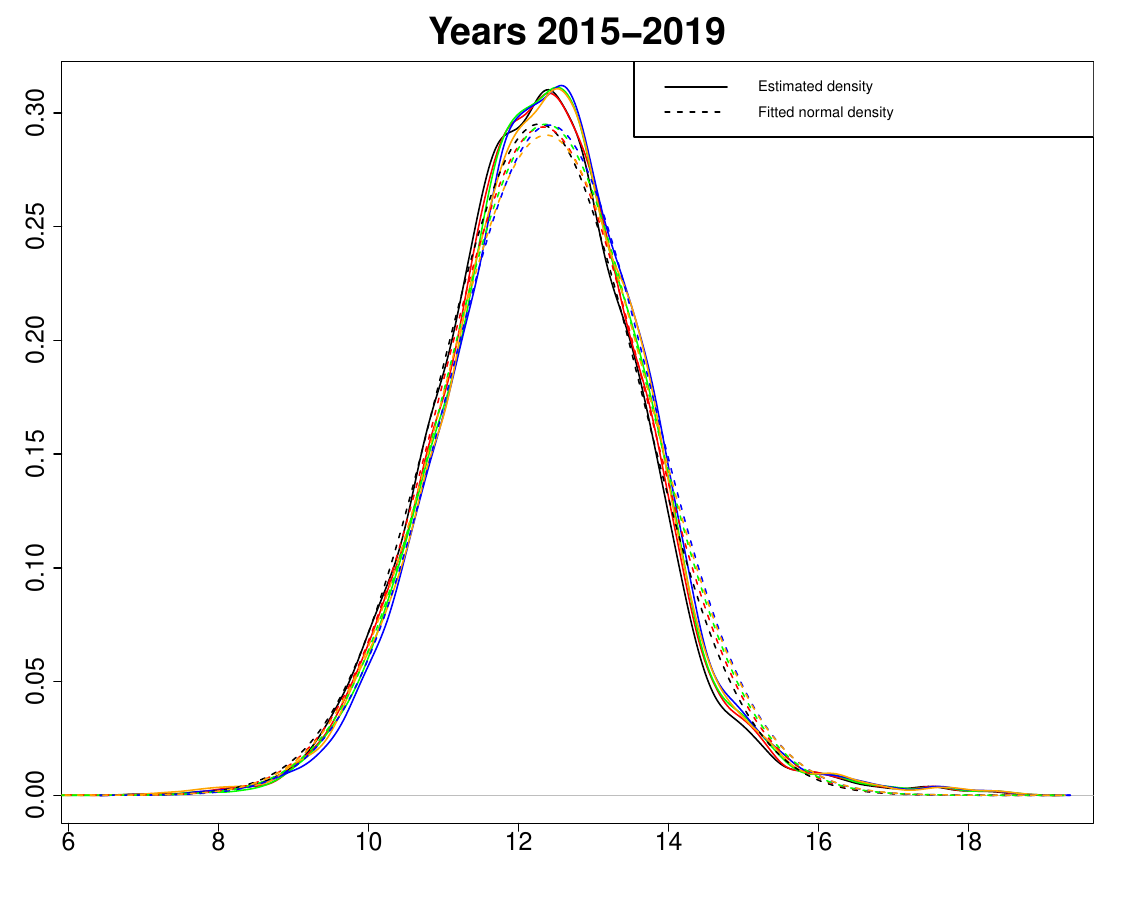} 
\caption{Kernel density estimators based on the log-transformed revenue data for 2015--2019, using the Gaussian kernel and Silverman's rule-of-thumb bandwidth \citep{silverman1986density}. 
Each pair of solid and dashed curves depicts the estimated density function and fitted normal density function respectively for one year.} 
\label{realdata_kde}
\end{figure}

In this section, we study the efficiency of the DRM-based quantile estimator and its parametric and nonparametric competitors with a real-world data. 
We use the collegiate sports budgets dataset from the TidyTuesday data project \citep{tidytuesday}, which is accessible from the Github repository (\url {https://github.com/rfordatascience/tidytuesday/tree/master/data/2022/2022-03-29}). 
The dataset contains yearly samples concerning some demographics of collegiate sports in the U.S. from 2015 to 2019.
The variable we consider is the total revenue (in USD) per sports team for both men and women, for which we have approximately 17,000 observations each year, and we log-transform the values to make the scale more suitable for numerical computation. 
As an exploratory data analysis, we plot in Figure~\ref {realdata_hist} the histograms of the log-transformed revenue data for the years 2015--2019. 
The population distributions of the revenues in these years look similar. 
This is also reflected in their kernel density estimators \citep{silverman1986density} as depicted in the solid curves in Figure~\ref {realdata_kde}. 
On one hand, the density estimates are close to bell-shaped, suggesting quantile estimation based on normal model is bearable. 
On the other hand, the estimated densities sufficiently deviate normal density as they have two or more modes, which can be better illustrated by comparing the estimated densities with the fitted normal densities depicted in the dashed curves. 
However, the estimated densities appear to have some common structures. 
Therefore, a DRM-based approach is more convincing and may work well in this situation. 

We conduct a real data-based simulation as follows. 
We regard the yearly samples from 2015--2019 as five populations, and sample with replacement from these populations to form multiple samples repeatedly. 
As remarked previously, our two-sample results are applicable to the multi-sample situation when $n_0/n_j \to \infty$ for all $j \neq 0$. 
We hence make the sample size from 2015 substantially larger to mimic the situation of $n_0/n_j$ being very large. 
This is meaningful as we can see how a large historical dataset could help predict the future with small datasets under the DRM. 
Specifically, we sample from 2016--2019 with equal sizes $ n \in \{ 200, 500 \} $, and sample from 2015 with size $ n_0 \in \{ 200, 1000, 5000 \} $.  
To apply the DRM-based approach, the user needs to specify a basis function $ \bq (x) $.
In this simulation, we use
i) data-adaptive basis function $ \bq (x) $ \citep{zhang2022density} learned using the full data with $d=2$; 
ii) $ \bq (x) = (1, x)^{\top} $ whose DRM contains the normal model with equal variances; 
iii) $ \bq (x) = (1, x, x^2)^{\top} $ whose DRM contains the normal model without equal variance assumption.
The simulation also correspondingly includes
iv) the parametric MLE of quantile derived under the normal model with common variance assumption; 
v) the parametric MLE under the normal model with no assumption on equal variances; and 
vi) the empirical quantiles.
By regarding the empirical quantiles based on the full data (treated as the populations) as the truth, we compute the simulated absolute biases, variances, and mean squared errors (MSEs) of these quantile estimators at some selected levels $ p $ for each population from 2016 to 2019. 
To save space, we report only the average values of the aforementioned performance measures across 2016--2019 in Tables~\ref {eff_quan_real2}--\ref {eff_quan_real3}.

We first observe that when $ n $ is fixed at 200 and 500 while $ n_0 $ increases from 200 to 5000, the variances of the DRM-based quantile estimators with various $ \bq (x) $ decrease in nearly all the cases. 
This observation supports our theoretical results. 
Second, the DRM-based quantile estimators are more efficient than the nonparametric empirical quantiles, which suggests that data pooling via the DRM works well for this data. 
Third, when $ n_0 = 5000 $, the quantile estimators derived under the DRM with $ \bq (x) = (1, x)^{\top} $ overall have comparable MSEs with the parametric estimators under the normal model with common variance assumption, except for the $90\%$ quantile level. 
The same also applies to the comparison between the estimators under DRM with $ \bq (x) = (1, x, x^2)^{\top} $ and the parametric estimators under the normal model with no equal variance assumption. 
This phenomenon can partly be explained by seeing in Figure~\ref {realdata_kde} that the true distributions deviate noticeably from the fitted normal distributions at the the right tails, which suggests normal model is unsatisfactory there. 
Finally, the DRM quantile estimators using the data-adaptive basis function $ \bq (x) $ generally beat the DRM estimators using the two prespecified $ \bq (x) $ in all three performance measures, except for very few cases where the results are still comparable. 
This is expected because when the basis function $ \bq (x) $ is adaptively learned from the data, the resulting DRM should fit the data better than a predetermined DRM. 
In fact, when $ n_0 = 5000 $, the adaptive DRM produces overall the most accurate quantile estimators, indicating that an appropriate DRM is suitable to the data. 

\begin{table}[h!] 
\begingroup
\setlength{\tabcolsep}{7pt} % Default value: 6pt
\renewcommand{\arraystretch}{0.8} % Default value: 1
\caption{
Simulated average absolute bias ($ \times \sqrt {n} $), variance ($ \times n $), and MSE ($ \times n $)
based on multiple samples from the real data. }
\label{eff_quan_real2}
\begin{center}
\begin{adjustbox}{max width=\textwidth}
\begin{tabular}{lrrrrrrrrrrrrrr}
\hline
\multicolumn{1}{l}{Level $ p $}         
& \multicolumn{3}{c}{DRM (adaptive $ \bq (x) $)} 
& \multicolumn{3}{c}{DRM ($ \bq (x) = (1, x)^{\top} $)} 
& \multicolumn{3}{c}{MLE (common-var normal)}  
& \multicolumn{3}{c}{Empirical quantile}  
\\ 
\cmidrule(lr){2-4} \cmidrule(lr){5-7} \cmidrule(lr){8-10} \cmidrule(lr){11-13} 
& Abs Bias & Var & MSE & Abs Bias & Var & MSE & Abs Bias & Var & MSE & Abs Bias & Var & MSE \\ 
\hline
& \multicolumn{12}{c}{$ n_0 = 200, n = 200 $} \\ 
\cline{2-13}
0.01      &  
$ 2.39 $ & $ 9.79 $ & $ 10.79 $ & $ 2.35 $ & $ 8.02 $ & $ 8.94 $ & $ 1.71 $ & $ 3.18 $ & $ 4.50 $ & $ 4.31 $ & $ 26.16 $ & $ 28.99 $ 
\\    
0.05      &  
$ 1.21 $ & $ 1.78 $ & $ 2.33 $ & $ 1.41 $ & $ 3.02 $ & $ 3.14 $ & $ 1.45 $ & $ 2.44 $ & $ 3.23 $ & $ 2.27 $ & $ 7.77 $ & $ 7.93 $ 
\\   
0.10      &  
$ 0.98 $ & $ 1.16 $ & $ 1.54 $ & $ 1.24 $ & $ 2.41 $ & $ 2.44 $ & $ 1.36 $ & $ 2.15 $ & $ 2.87 $ & $ 1.81 $ & $ 5.15 $ & $ 5.16 $ 
\\   
0.50      &  
$ 0.76 $ & $ 0.60 $ & $ 0.90 $ & $ 1.04 $ & $ 1.70 $ & $ 1.71 $ & $ 1.06 $ & $ 1.80 $ & $ 1.81 $ & $ 1.25 $ & $ 2.44 $ & $ 2.45 $ 
\\   
0.90      &  
$ 0.93 $ & $ 1.10 $ & $ 1.34 $ & $ 1.20 $ & $ 2.27 $ & $ 2.28 $ & $ 1.93 $ & $ 2.45 $ & $ 5.38 $ & $ 1.74 $ & $ 4.70 $ & $ 4.75 $ 
\\   
0.95      &  
$ 1.63 $ & $ 3.59 $ & $ 4.06 $ & $ 2.02 $ & $ 6.34 $ & $ 6.40 $ & $ 1.64 $ & $ 2.82 $ & $ 4.12 $ & $ 3.03 $ & $ 13.59 $ & $ 13.87 $ 
\\   
& \multicolumn{12}{c}{$ n_0 = 1000, n = 200 $} \\ 
\cline{2-13}
0.01      &  
$ 1.80 $ & $ 5.12 $ & $ 6.10 $ & $ 1.91 $ & $ 4.74 $ & $ 5.84 $ & $ 1.55 $ & $ 2.57 $ & $ 3.71 $ & $ 4.27 $ & $ 25.61 $ & $ 28.78 $ 
\\    
0.05      &  
$ 1.09 $ & $ 0.90 $ & $ 1.88 $ & $ 1.24 $ & $ 2.36 $ & $ 2.46 $ & $ 1.34 $ & $ 2.21 $ & $ 2.84 $ & $ 2.29 $ & $ 7.78 $ & $ 8.02 $ 
\\   
0.10      &  
$ 0.98 $ & $ 0.67 $ & $ 1.52 $ & $ 1.13 $ & $ 2.03 $ & $ 2.05 $ & $ 1.30 $ & $ 2.07 $ & $ 2.66 $ & $ 1.80 $ & $ 5.03 $ & $ 5.05 $ 
\\   
0.50      &  
$ 0.82 $ & $ 0.40 $ & $ 1.02 $ & $ 1.03 $ & $ 1.69 $ & $ 1.71 $ & $ 1.10 $ & $ 1.90 $ & $ 1.92 $ & $ 1.29 $ & $ 2.60 $ & $ 2.61 $ 
\\   
0.90      &  
$ 0.91 $ & $ 0.73 $ & $ 1.23 $ & $ 1.18 $ & $ 2.20 $ & $ 2.22 $ & $ 1.86 $ & $ 2.24 $ & $ 5.03 $ & $ 1.75 $ & $ 4.71 $ & $ 4.75 $ 
\\   
0.95      &  
$ 1.38 $ & $ 2.10 $ & $ 2.89 $ & $ 1.93 $ & $ 5.83 $ & $ 5.94 $ & $ 1.52 $ & $ 2.43 $ & $ 3.60 $ & $ 2.97 $ & $ 13.14 $ & $ 13.32 $ 
\\   
& \multicolumn{12}{c}{$ n_0 = 5000, n = 200 $} \\ 
\cline{2-13}
0.01      &  
$ 1.39 $ & $ 2.88 $ & $ 3.93 $ & $ 1.62 $ & $ 2.39 $ & $ 4.12 $ & $ 1.41 $ & $ 2.02 $ & $ 3.05 $ & $ 4.29 $ & $ 25.14 $ & $ 28.56 $ 
\\    
0.05      &  
$ 1.15 $ & $ 0.39 $ & $ 1.97 $ & $ 1.09 $ & $ 1.77 $ & $ 1.90 $ & $ 1.25 $ & $ 1.90 $ & $ 2.42 $ & $ 2.24 $ & $ 7.49 $ & $ 7.74 $ 
\\   
0.10      &  
$ 1.18 $ & $ 0.29 $ & $ 1.85 $ & $ 1.01 $ & $ 1.61 $ & $ 1.63 $ & $ 1.23 $ & $ 1.86 $ & $ 2.33 $ & $ 1.76 $ & $ 4.78 $ & $ 4.81 $ 
\\   
0.50      &  
$ 0.99 $ & $ 0.19 $ & $ 1.32 $ & $ 0.98 $ & $ 1.50 $ & $ 1.52 $ & $ 1.08 $ & $ 1.80 $ & $ 1.81 $ & $ 1.28 $ & $ 2.55 $ & $ 2.55 $ 
\\   
0.90      &  
$ 0.94 $ & $ 0.34 $ & $ 1.25 $ & $ 1.08 $ & $ 1.84 $ & $ 1.84 $ & $ 1.79 $ & $ 1.88 $ & $ 4.60 $ & $ 1.72 $ & $ 4.53 $ & $ 4.57 $ 
\\   
0.95      &  
$ 1.28 $ & $ 0.91 $ & $ 2.35 $ & $ 1.80 $ & $ 5.02 $ & $ 5.18 $ & $ 1.40 $ & $ 1.93 $ & $ 3.03 $ & $ 2.99 $ & $ 13.22 $ & $ 13.42 $ 
\\   
& \multicolumn{12}{c}{$ n_0 = 200, n = 500 $} \\ 
\cline{2-13}
0.01      &  
$ 2.53 $ & $ 8.50 $ & $ 10.71 $ & $ 2.63 $ & $ 8.74 $ & $ 10.82 $ & $ 2.08 $ & $ 3.40 $ & $ 6.47 $ & $ 4.35 $ & $ 28.77 $ & $ 30.34 $ 
\\    
0.05      &  
$ 1.42 $ & $ 1.97 $ & $ 3.12 $ & $ 1.50 $ & $ 3.27 $ & $ 3.50 $ & $ 1.72 $ & $ 2.58 $ & $ 4.46 $ & $ 2.31 $ & $ 7.87 $ & $ 7.97 $ 
\\   
0.10      &  
$ 1.12 $ & $ 1.32 $ & $ 1.99 $ & $ 1.27 $ & $ 2.55 $ & $ 2.58 $ & $ 1.63 $ & $ 2.27 $ & $ 3.97 $ & $ 1.85 $ & $ 5.34 $ & $ 5.34 $ 
\\   
0.50      &  
$ 0.92 $ & $ 0.68 $ & $ 1.31 $ & $ 1.07 $ & $ 1.81 $ & $ 1.85 $ & $ 1.09 $ & $ 1.86 $ & $ 1.90 $ & $ 1.28 $ & $ 2.60 $ & $ 2.61 $ 
\\   
0.90      &  
$ 1.04 $ & $ 1.22 $ & $ 1.69 $ & $ 1.24 $ & $ 2.40 $ & $ 2.41 $ & $ 2.81 $ & $ 2.53 $ & $ 10.14 $ & $ 1.74 $ & $ 4.65 $ & $ 4.67 $ 
\\   
0.95      &  
$ 1.69 $ & $ 3.62 $ & $ 4.49 $ & $ 2.06 $ & $ 6.58 $ & $ 6.74 $ & $ 2.10 $ & $ 2.92 $ & $ 6.39 $ & $ 3.03 $ & $ 14.08 $ & $ 14.23 $ 
\\   
& \multicolumn{12}{c}{$ n_0 = 1000, n = 500 $} \\ 
\cline{2-13}
0.01      &  
$ 2.17 $ & $ 5.60 $ & $ 7.75 $ & $ 2.36 $ & $ 6.28 $ & $ 8.82 $ & $ 1.99 $ & $ 2.99 $ & $ 5.91 $ & $ 4.46 $ & $ 29.63 $ & $ 31.31 $ 
\\    
0.05      &  
$ 1.34 $ & $ 1.33 $ & $ 2.86 $ & $ 1.40 $ & $ 2.79 $ & $ 3.05 $ & $ 1.62 $ & $ 2.41 $ & $ 4.03 $ & $ 2.33 $ & $ 8.17 $ & $ 8.32 $ 
\\   
0.10      &  
$ 1.13 $ & $ 0.96 $ & $ 2.03 $ & $ 1.21 $ & $ 2.29 $ & $ 2.31 $ & $ 1.55 $ & $ 2.19 $ & $ 3.67 $ & $ 1.84 $ & $ 5.28 $ & $ 5.29 $ 
\\   
0.50      &  
$ 0.98 $ & $ 0.52 $ & $ 1.46 $ & $ 1.07 $ & $ 1.75 $ & $ 1.77 $ & $ 1.12 $ & $ 1.90 $ & $ 1.93 $ & $ 1.32 $ & $ 2.69 $ & $ 2.70 $ 
\\   
0.90      &  
$ 1.06 $ & $ 0.97 $ & $ 1.73 $ & $ 1.21 $ & $ 2.24 $ & $ 2.27 $ & $ 2.74 $ & $ 2.37 $ & $ 9.64 $ & $ 1.72 $ & $ 4.56 $ & $ 4.60 $ 
\\   
0.95      &  
$ 1.67 $ & $ 2.91 $ & $ 4.31 $ & $ 2.03 $ & $ 6.28 $ & $ 6.48 $ & $ 2.00 $ & $ 2.65 $ & $ 5.80 $ & $ 3.03 $ & $ 13.95 $ & $ 14.12 $ 
\\   
& \multicolumn{12}{c}{$ n_0 = 5000, n = 500 $} \\ 
\cline{2-13}
0.01      &  
$ 1.73 $ & $ 2.73 $ & $ 4.92 $ & $ 2.09 $ & $ 3.27 $ & $ 7.06 $ & $ 1.79 $ & $ 2.28 $ & $ 4.82 $ & $ 4.44 $ & $ 30.08 $ & $ 31.55 $ 
\\    
0.05      &  
$ 1.47 $ & $ 0.65 $ & $ 3.39 $ & $ 1.19 $ & $ 1.98 $ & $ 2.27 $ & $ 1.49 $ & $ 2.03 $ & $ 3.35 $ & $ 2.29 $ & $ 7.77 $ & $ 7.90 $ 
\\   
0.10      &  
$ 1.46 $ & $ 0.48 $ & $ 3.02 $ & $ 1.06 $ & $ 1.76 $ & $ 1.79 $ & $ 1.44 $ & $ 1.94 $ & $ 3.15 $ & $ 1.78 $ & $ 4.94 $ & $ 4.96 $ 
\\   
0.50      &  
$ 1.32 $ & $ 0.29 $ & $ 2.42 $ & $ 1.00 $ & $ 1.55 $ & $ 1.59 $ & $ 1.08 $ & $ 1.80 $ & $ 1.86 $ & $ 1.28 $ & $ 2.60 $ & $ 2.60 $ 
\\   
0.90      &  
$ 1.24 $ & $ 0.50 $ & $ 2.21 $ & $ 1.12 $ & $ 1.96 $ & $ 1.96 $ & $ 2.70 $ & $ 1.98 $ & $ 9.09 $ & $ 1.73 $ & $ 4.57 $ & $ 4.58 $ 
\\   
0.95      &  
$ 1.68 $ & $ 1.43 $ & $ 4.23 $ & $ 1.86 $ & $ 5.48 $ & $ 5.77 $ & $ 1.88 $ & $ 2.09 $ & $ 5.08 $ & $ 3.10 $ & $ 14.60 $ & $ 14.66 $ 
\\   
\hline
\end{tabular}
\end{adjustbox}
\end{center}
\endgroup
\end{table}

\begin{table}[h!] 
\begingroup
\setlength{\tabcolsep}{7pt} % Default value: 6pt
\renewcommand{\arraystretch}{0.8} % Default value: 1
\caption{
Simulated average absolute bias ($ \times \sqrt {n} $), variance ($ \times n $), and MSE ($ \times n $)
based on multiple samples from the real data. }
\label{eff_quan_real3}
\begin{center}
\begin{adjustbox}{max width=\textwidth}
\begin{tabular}{lrrrrrrrrrrrrrr}
\hline
\multicolumn{1}{l}{Level $ p $}         
& \multicolumn{3}{c}{DRM (adaptive $ \bq (x) $)} 
& \multicolumn{3}{c}{DRM ($ \bq (x) = (1, x, x^2)^{\top} $)} 
& \multicolumn{3}{c}{MLE (normal)}  
& \multicolumn{3}{c}{Empirical quantile}  
\\ 
\cmidrule(lr){2-4} \cmidrule(lr){5-7} \cmidrule(lr){8-10} \cmidrule(lr){11-13} 
& Abs Bias & Var & MSE & Abs Bias & Var & MSE & Abs Bias & Var & MSE & Abs Bias & Var & MSE \\ 
\hline
& \multicolumn{12}{c}{$ n_0 = 200, n = 200 $} \\ 
\cline{2-13}
0.01      &  
$ 2.39 $ & $ 9.79 $ & $ 10.79 $ & $ 2.92 $ & $ 13.90 $ & $ 14.42 $ & $ 2.40 $ & $ 8.47 $ & $ 9.16 $ & $ 4.31 $ & $ 26.16 $ & $ 28.99 $ 
\\    
0.05      &  
$ 1.21 $ & $ 1.78 $ & $ 2.33 $ & $ 1.76 $ & $ 4.82 $ & $ 4.83 $ & $ 1.86 $ & $ 4.90 $ & $ 5.47 $ & $ 2.27 $ & $ 7.77 $ & $ 7.93 $ 
\\   
0.10      &  
$ 0.98 $ & $ 1.16 $ & $ 1.54 $ & $ 1.48 $ & $ 3.45 $ & $ 3.48 $ & $ 1.62 $ & $ 3.54 $ & $ 4.16 $ & $ 1.81 $ & $ 5.15 $ & $ 5.16 $ 
\\   
0.50      &  
$ 0.76 $ & $ 0.60 $ & $ 0.90 $ & $ 1.05 $ & $ 1.74 $ & $ 1.75 $ & $ 1.06 $ & $ 1.80 $ & $ 1.81 $ & $ 1.25 $ & $ 2.44 $ & $ 2.45 $ 
\\   
0.90      &  
$ 0.93 $ & $ 1.10 $ & $ 1.34 $ & $ 1.42 $ & $ 3.17 $ & $ 3.20 $ & $ 2.20 $ & $ 4.80 $ & $ 7.57 $ & $ 1.74 $ & $ 4.70 $ & $ 4.75 $ 
\\   
0.95      &  
$ 1.63 $ & $ 3.59 $ & $ 4.06 $ & $ 2.53 $ & $ 9.92 $ & $ 9.99 $ & $ 2.19 $ & $ 6.51 $ & $ 7.69 $ & $ 3.03 $ & $ 13.59 $ & $ 13.87 $ 
\\   
& \multicolumn{12}{c}{$ n_0 = 1000, n = 200 $} \\ 
\cline{2-13}
0.01      &  
$ 1.80 $ & $ 5.12 $ & $ 6.10 $ & $ 2.57 $ & $ 9.53 $ & $ 10.52 $ & $ 2.38 $ & $ 8.18 $ & $ 8.81 $ & $ 4.27 $ & $ 25.61 $ & $ 28.78 $ 
\\    
0.05      &  
$ 1.09 $ & $ 0.90 $ & $ 1.88 $ & $ 1.65 $ & $ 4.26 $ & $ 4.29 $ & $ 1.83 $ & $ 4.79 $ & $ 5.31 $ & $ 2.29 $ & $ 7.78 $ & $ 8.02 $ 
\\   
0.10      &  
$ 0.98 $ & $ 0.67 $ & $ 1.52 $ & $ 1.42 $ & $ 3.22 $ & $ 3.23 $ & $ 1.60 $ & $ 3.51 $ & $ 4.08 $ & $ 1.80 $ & $ 5.03 $ & $ 5.05 $ 
\\   
0.50      &  
$ 0.82 $ & $ 0.40 $ & $ 1.02 $ & $ 1.04 $ & $ 1.73 $ & $ 1.75 $ & $ 1.10 $ & $ 1.90 $ & $ 1.92 $ & $ 1.29 $ & $ 2.60 $ & $ 2.61 $ 
\\   
0.90      &  
$ 0.91 $ & $ 0.73 $ & $ 1.23 $ & $ 1.41 $ & $ 3.07 $ & $ 3.10 $ & $ 2.20 $ & $ 4.83 $ & $ 7.56 $ & $ 1.75 $ & $ 4.71 $ & $ 4.75 $ 
\\   
0.95      &  
$ 1.38 $ & $ 2.10 $ & $ 2.89 $ & $ 2.52 $ & $ 9.73 $ & $ 9.79 $ & $ 2.20 $ & $ 6.49 $ & $ 7.61 $ & $ 2.97 $ & $ 13.14 $ & $ 13.32 $ 
\\   
& \multicolumn{12}{c}{$ n_0 = 5000, n = 200 $} \\ 
\cline{2-13}
0.01      &  
$ 1.39 $ & $ 2.88 $ & $ 3.93 $ & $ 2.32 $ & $ 6.35 $ & $ 8.37 $ & $ 2.36 $ & $ 8.06 $ & $ 8.77 $ & $ 4.29 $ & $ 25.14 $ & $ 28.56 $ 
\\    
0.05      &  
$ 1.15 $ & $ 0.39 $ & $ 1.97 $ & $ 1.58 $ & $ 3.86 $ & $ 3.96 $ & $ 1.81 $ & $ 4.71 $ & $ 5.20 $ & $ 2.24 $ & $ 7.49 $ & $ 7.74 $ 
\\   
0.10      &  
$ 1.18 $ & $ 0.29 $ & $ 1.85 $ & $ 1.35 $ & $ 2.93 $ & $ 2.95 $ & $ 1.58 $ & $ 3.43 $ & $ 3.95 $ & $ 1.76 $ & $ 4.78 $ & $ 4.81 $ 
\\   
0.50      &  
$ 0.99 $ & $ 0.19 $ & $ 1.32 $ & $ 1.00 $ & $ 1.55 $ & $ 1.57 $ & $ 1.08 $ & $ 1.80 $ & $ 1.81 $ & $ 1.28 $ & $ 2.55 $ & $ 2.55 $ 
\\   
0.90      &  
$ 0.94 $ & $ 0.34 $ & $ 1.25 $ & $ 1.32 $ & $ 2.70 $ & $ 2.72 $ & $ 2.19 $ & $ 4.60 $ & $ 7.47 $ & $ 1.72 $ & $ 4.53 $ & $ 4.57 $ 
\\   
0.95      &  
$ 1.28 $ & $ 0.91 $ & $ 2.35 $ & $ 2.41 $ & $ 9.07 $ & $ 9.12 $ & $ 2.15 $ & $ 6.21 $ & $ 7.43 $ & $ 2.99 $ & $ 13.22 $ & $ 13.42 $ 
\\   
& \multicolumn{12}{c}{$ n_0 = 200, n = 500 $} \\ 
\cline{2-13}
0.01      &  
$ 2.53 $ & $ 8.50 $ & $ 10.71 $ & $ 3.03 $ & $ 13.27 $ & $ 14.59 $ & $ 2.53 $ & $ 8.17 $ & $ 10.08 $ & $ 4.35 $ & $ 28.77 $ & $ 30.34 $ 
\\    
0.05      &  
$ 1.42 $ & $ 1.97 $ & $ 3.12 $ & $ 1.79 $ & $ 4.86 $ & $ 4.90 $ & $ 2.03 $ & $ 4.81 $ & $ 6.38 $ & $ 2.31 $ & $ 7.87 $ & $ 7.97 $ 
\\   
0.10      &  
$ 1.12 $ & $ 1.32 $ & $ 1.99 $ & $ 1.51 $ & $ 3.57 $ & $ 3.60 $ & $ 1.83 $ & $ 3.53 $ & $ 5.17 $ & $ 1.85 $ & $ 5.34 $ & $ 5.34 $ 
\\   
0.50      &  
$ 0.92 $ & $ 0.68 $ & $ 1.31 $ & $ 1.08 $ & $ 1.84 $ & $ 1.88 $ & $ 1.09 $ & $ 1.86 $ & $ 1.90 $ & $ 1.28 $ & $ 2.60 $ & $ 2.61 $ 
\\   
0.90      &  
$ 1.04 $ & $ 1.22 $ & $ 1.69 $ & $ 1.41 $ & $ 3.07 $ & $ 3.09 $ & $ 2.92 $ & $ 4.61 $ & $ 12.07 $ & $ 1.74 $ & $ 4.65 $ & $ 4.67 $ 
\\   
0.95      &  
$ 1.69 $ & $ 3.62 $ & $ 4.49 $ & $ 2.50 $ & $ 9.66 $ & $ 9.72 $ & $ 2.48 $ & $ 6.20 $ & $ 9.52 $ & $ 3.03 $ & $ 14.08 $ & $ 14.23 $ 
\\   
& \multicolumn{12}{c}{$ n_0 = 1000, n = 500 $} \\ 
\cline{2-13}
0.01      &  
$ 2.17 $ & $ 5.60 $ & $ 7.75 $ & $ 2.86 $ & $ 10.72 $ & $ 12.83 $ & $ 2.57 $ & $ 8.24 $ & $ 10.29 $ & $ 4.46 $ & $ 29.63 $ & $ 31.31 $ 
\\    
0.05      &  
$ 1.34 $ & $ 1.33 $ & $ 2.86 $ & $ 1.72 $ & $ 4.53 $ & $ 4.62 $ & $ 2.02 $ & $ 4.87 $ & $ 6.31 $ & $ 2.33 $ & $ 8.17 $ & $ 8.32 $ 
\\   
0.10      &  
$ 1.13 $ & $ 0.96 $ & $ 2.03 $ & $ 1.47 $ & $ 3.40 $ & $ 3.41 $ & $ 1.81 $ & $ 3.58 $ & $ 5.06 $ & $ 1.84 $ & $ 5.28 $ & $ 5.29 $ 
\\   
0.50      &  
$ 0.98 $ & $ 0.52 $ & $ 1.46 $ & $ 1.08 $ & $ 1.78 $ & $ 1.81 $ & $ 1.12 $ & $ 1.90 $ & $ 1.93 $ & $ 1.32 $ & $ 2.69 $ & $ 2.70 $ 
\\   
0.90      &  
$ 1.06 $ & $ 0.97 $ & $ 1.73 $ & $ 1.39 $ & $ 2.96 $ & $ 2.99 $ & $ 2.90 $ & $ 4.65 $ & $ 11.94 $ & $ 1.72 $ & $ 4.56 $ & $ 4.60 $ 
\\   
0.95      &  
$ 1.67 $ & $ 2.91 $ & $ 4.31 $ & $ 2.50 $ & $ 9.67 $ & $ 9.76 $ & $ 2.47 $ & $ 6.23 $ & $ 9.39 $ & $ 3.03 $ & $ 13.95 $ & $ 14.12 $ 
\\   
& \multicolumn{12}{c}{$ n_0 = 5000, n = 500 $} \\ 
\cline{2-13}
0.01      &  
$ 1.73 $ & $ 2.73 $ & $ 4.92 $ & $ 2.66 $ & $ 7.47 $ & $ 11.04 $ & $ 2.54 $ & $ 8.33 $ & $ 10.08 $ & $ 4.44 $ & $ 30.08 $ & $ 31.55 $ 
\\    
0.05      &  
$ 1.47 $ & $ 0.65 $ & $ 3.39 $ & $ 1.59 $ & $ 3.98 $ & $ 4.07 $ & $ 2.00 $ & $ 4.84 $ & $ 6.28 $ & $ 2.29 $ & $ 7.77 $ & $ 7.90 $ 
\\   
0.10      &  
$ 1.46 $ & $ 0.48 $ & $ 3.02 $ & $ 1.37 $ & $ 2.98 $ & $ 3.01 $ & $ 1.80 $ & $ 3.51 $ & $ 5.02 $ & $ 1.78 $ & $ 4.94 $ & $ 4.96 $ 
\\   
0.50      &  
$ 1.32 $ & $ 0.29 $ & $ 2.42 $ & $ 1.01 $ & $ 1.59 $ & $ 1.63 $ & $ 1.08 $ & $ 1.80 $ & $ 1.86 $ & $ 1.28 $ & $ 2.60 $ & $ 2.60 $ 
\\   
0.90      &  
$ 1.24 $ & $ 0.50 $ & $ 2.21 $ & $ 1.35 $ & $ 2.79 $ & $ 2.82 $ & $ 2.98 $ & $ 4.72 $ & $ 12.55 $ & $ 1.73 $ & $ 4.57 $ & $ 4.58 $ 
\\   
0.95      &  
$ 1.68 $ & $ 1.43 $ & $ 4.23 $ & $ 2.52 $ & $ 9.90 $ & $ 10.06 $ & $ 2.55 $ & $ 6.39 $ & $ 10.00 $ & $ 3.10 $ & $ 14.60 $ & $ 14.66 $ 
\\   
\hline
\end{tabular}
\end{adjustbox}
\end{center}
\endgroup
\end{table}

%%%%%%%%%%%%%%%%%%%%%%%%%%%%%%%%%%%%%%%%%%%%%%%%%%%%%%%%%%%%%%%%%%%%%%%%%%%%%%%%%%%%%%%%%%%%%%%%%%%%%%%%%%%%%%%%%%%%%%%%%%%%%%
 
\section{Conclusions and discussion} 
\label{sec:conclusion} 

The DRM for multi-sample data is generating growing research interest and has wide applications in statistics and econometrics. 
It has proven particularly useful in situations where the populations may possess shared underlying structures. 
The DRM provides a good trade-off between low model misspecification risk and satisfactory statistical efficiency. 
Its effectiveness primarily stems from its ability to enable users to draw inferences on each population using pooled data, which leads to efficiency gain compared to using individual samples that overlooks the shared latent structures. 
The literature has engaged in discussions regarding the efficiency of the DRM approach in comparison to the nonparametric approach. However, none of these discussions have systematically explored the limit of the efficiency gain through the DRM, which is an important research problem. 
This article addresses this problem by considering a scenario where one of the samples significantly outweighs the others in size. 
We establish through theoretical analysis that within this context, the DRM-based estimators of model parameters, distribution functions, and quantiles for the smaller sample populations attain the efficiency as if a parametric model is assumed. 
In essence, for these estimands and in this scenario, we identify their highest achievable efficiency under a specific parametric model, investigate their asymptotic efficiency under the DRM, and demonstrate the equivalence between the two aforementioned efficiencies. 
Our simulation experiments and analyses of real-world data, with a particular focus on quantile estimation, support our theoretical discoveries.
The significance of this article's contribution extends to practical scenarios where researchers aim to make inferences about a population with limited data, but can rely on a substantial dataset from a related population for support. 

While the scenario examined in this article addresses many real-world applications, there exist situations where this sampling scheme is not applicable. 
In such cases, we may encounter a growing number of samples, each of relatively similar sizes. 
For example, when economists investigate the evolution of income distribution over time, they may collect income samples year after year, and these yearly samples often have comparable sizes. 
Consequently, it becomes intriguing to investigate the asymptotic efficiency of the DRM approach when the sample sizes $ n_i/n_j \to 1 $ and the number of populations $ m \to \infty $ as $ n_i \to \infty $.  
Additionally, it is worthwhile to study the asymptotic efficiency of other population parameters than the ones we investigated. 
We leave these interesting problems for future work, anticipating that the technical methods and theoretical results in this article may be valuable for such inquiries. 

\section*{Acknowledgement} 
This research was partially supported by the Natural Sciences and Engineering Research Council of Canada (Grants RGPIN-2018-06484, RGPIN-2019-04204, and RGPIN-2020-05897), the Canadian Statistical Sciences Institute (Grant 592307), and the Department of Statistical Sciences in the University of Toronto. 
We also thank the Digital Research Alliance of Canada for computing support. 
This work was partially completed when Archer Gong Zhang was a PhD student at the University of British Columbia and a Postdoctoral Fellow at the University of Toronto.
\clearpage
\appendix
\section{Appendix} 
\label{sec:appendix}

In this section, we provide the proofs of the theoretical results in Section~\ref{sec:EL-DRM}.  

\subsection{Proof of Lemma~\ref{lem:profile}} 

This lemma claims some properties of the profile log-EL $ \ell_{n} (\btheta) $ and its derivatives. 

\begin{proof}

We start with the first conclusion of the lemma that the score function $ \partial \ell_{n} (\btheta)/\partial \btheta $ has expectation zero. 
Let $ \rho_{n, 0} = n_0/n, \, \rho_{n, 1} = n_1/n $ denote the sample proportions, where $ n = n_0 + n_1 $ is the total sample size.  
By Condition~\eqref{Condition.i}, $ \rho_{n, 0} \to 1, \, \rho_{n, 1} \to 0 $ as $ n_0, n_1 \to \infty $. 
With this notation, the profile log-EL in \eqref{dual-log-EL} can be written as 
\begin{align*} 
\ell_{n} (\btheta) = 
- \sum_{k, j} \log \left [ \rho_{n, 0} + \rho_{n, 1} \exp \left \{ \btheta^{\top} \bq (x_{k j}) \right \} \right ] 
+ \sum_{j = 1}^{n_1} \btheta^{\top} \bq (x_{1 j}) 
- n \log n. 
\end{align*}
The EL-based score function is then
\[ 
\frac {\partial \ell_{n} (\btheta)} {\partial \btheta} 
= 
- \sum_{k, j} \frac {\rho_{n, 1} \exp \{ \btheta^{\top} \bq (x_{k j}) \} \bq (x_{k j})} {\rho_{n, 0} + \rho_{n, 1} \exp \{ \btheta^{\top} \bq (x_{k j}) \}}
+ \sum_{j = 1}^{n_1} \bq (x_{1 j}). 
\] 
For notational convenience, we let 
\[ 
h_n (x, \btheta) = \frac {\exp \{ \btheta^{\top} \bq (x) \}} {\rho_{n, 0} + \rho_{n, 1} \exp \{ \btheta^{\top} \bq (x) \}},
\] 
where the subscript $ n $ is to emphasize that it evolves with the sample size $ n $. 

It is then seen that at $ \btheta = \btheta^{*} $, we have 
\begin{align*} 
\bbE \Big [\frac {\partial \ell_n (\btheta^{*})} {\partial \btheta} \Big ] 
& = -\sum_k n_k \bbE_k [ \rho_{n, 1} h_n (X, \btheta^*) \bq (X)] + n_1 \bbE_1 [ \bq (X)] \\ 
& = -\bbE_0 [ \rho_{n, 1} h_n (X, \btheta^*) \bq (X) \{ n_0 + n_1 \exp (\btheta^{* \top} \bq (X)) \} ] + n_1 \bbE_1 [ \bq (X)] \\ 
& = -\bbE_0 [ n_1 \exp \{ \btheta^{* \top} \bq (X) \} \bq (X) ] + n_1 \bbE_1 [ \bq (X)] \\ 
& = \bsm{0}. 
\end{align*} 
This is the first conclusion of Lemma~\ref{lem:profile}.

The second conclusion is the asymptotic normality of the score function. 
Despite its complex expression, the score function is a sum of three sets of i.i.d. random variables: 
\begin{align} 
n_1^{-1/2} \frac {\partial \ell_n (\btheta^{*})} {\partial \btheta} 
= & n_1^{-1/2} \Big \{ \frac {\partial \ell_n (\btheta^{*})} {\partial \btheta} - \bbE \big [ \frac {\partial \ell_n (\btheta^{*})} {\partial \btheta} \big ] \Big \} \nonumber \\ 
= & n_1^{-1/2} \sum_{j=0}^{n_0} \rho_{n, 1} \{ -h_n (x_{0 j}, \btheta^*) \bq (x_{0 j}) - \bbE_0 [-h_n (X, \btheta^*) \bq (X)] \} \label{score_dec1} \\ 
+ & n_1^{-1/2} \sum_{j=1}^{n_1} \rho_{n, 1} \{ -h_n (x_{1 j}, \btheta^*) \bq (x_{1 j}) - \bbE_1 [-h_n (X, \btheta^*) \bq (X)] \} \label{score_dec2} \\ 
+ & n_1^{-1/2} \sum_{j=1}^{n_1} \{ \bq (x_{1 j}) - \bbE_1 [\bq (x_{1 j})] \} \label{score_dec3}. 
\end{align} 
We now study them term by term. 
Without loss of generality, we drop the expectations in the preceding terms \eqref{score_dec1}--\eqref{score_dec3} and proceed as if they are already centralized. 

The first term \eqref{score_dec1} can be written as 
\begin{align*} 
-n_1^{-1/2} \sum_{j=0}^{n_0} \rho_{n, 1} h_n (x_{0 j}, \btheta^*) \bq (x_{0 j}) 
= -(\rho_{n,0} \rho_{n,1})^{1/2} \Big \{ n_0^{-1/2} \sum_{j=0}^{n_0} h_n (x_{0 j}, \btheta^*) \bq (x_{0 j}) \Big \}. 
\end{align*} 
The term in the curly brackets is asymptotically normal with variance $ \Var_0 \left [\exp \{ \btheta^{* \top} \bq (X) \} \bq (X) \right ] $ by the central limit theorem for triangular arrays \citep[Theorem 3.4.10]{durrett2019probability} (also known as the Lindeberg--Feller central limit theorem). 
Note that the variance is finite by Condition~\eqref{Condition.iii}. 
Combined with that $ (\rho_{n,0} \rho_{n,1})^{1/2} = o (1) $, the term \eqref{score_dec1} is $ o_p (1) $. 

For the second term \eqref{score_dec2}, by Chebyshev's inequality, $ \forall \epsilon > 0 $, 
\[ 
P \left ( \Big \|-n_1^{-1/2} \sum_{j=1}^{n_1} \rho_{n, 1} h_n (x_{1 j}, \btheta^*) \bq (x_{1 j}) \Big \| \geq \epsilon \right ) 
\leq \epsilon^{-2} \bbE_1 \| \rho_{n,1} h_n (X, \btheta^*) \bq (X) \|^2. 
\] 
Since $ \| \rho_{n,1} h_n (x, \btheta^*) \bq (x) \| < \| \bq (x) \| $ for all $ x $ with $ \bbE_1 \| \bq (X) \|^2 < \infty $, and $ \rho_{n,1} h_n (x, \btheta^*) \bq (x) \to \bsm{0} $ pointwise, we have 
\[ 
\bbE_1 \| \rho_{n,1} h_n (X, \btheta^*) \bq (X) \|^2 \to 0, 
\] 
by the dominated convergence theorem. 
Therefore, the term \eqref{score_dec2} is $ o_p (1) $ by definition. 
{\it (Remark: we can not factor out $ \rho_{n,1} $ and apply the central limit theorem to the remaining part here because $ \Var_1 \left [\exp \{ \btheta^{* \top} \bq (X) \} \bq (X) \right ] $ may not be finite.)} 

The third term \eqref{score_dec3} is straightforward to study: by the central limit theorem, it has an asymptotic normal distribution with finite variance $ \Var_1 [\bq (X)] $ guaranteed by Condition~\eqref{Condition.iii}. 

Therefore, combining all the arguments above for the terms \eqref{score_dec1}--\eqref{score_dec3}, we arrive at the conclusion that as $ n_0, n_1 \to \infty $, 
\begin{align*} 
n_1^{-1/2} \frac {\partial \ell_n (\btheta^{*})} {\partial \btheta} 
= o_p (1) + o_p (1) + n_1^{-1/2} \sum_{j=1}^{n_1} \{ \bq (x_{1 j}) - \bbE_1 [\bq (x_{1 j})] \} 
\overset{d} \to N (\bsm{0}, \Var_1 [\bq (X)]), 
\end{align*} 
by Slutsky's theorem. 
This proves the second conclusion of Lemma~\ref{lem:profile}. 

Finally, the third conclusion of Lemma~\ref{lem:profile} states that the negative Hessian has a positive definite limit. 
At $ \btheta = \btheta^* $, the Hessian is the sum of two sets of i.i.d. random variables: 
\begin{align} 
-n_1^{-1} \frac {\partial^2 \ell_{n} (\btheta^{*})} {\partial \btheta \partial \btheta^{\top}} 
= & n_1^{-1} \sum_{j = 1}^{n_0} \bq (x_{0 j}) \bq^\top (x_{0 j}) \big \{ \rho_{n,1} h_n (x_{0 j}, \btheta^*) - \rho_{n,1}^2 h_n^2 (x_{0 j}, \btheta^*) \big \} \label{hessian_dec1} \\ 
& + n_1^{-1} \sum_{j = 1}^{n_1} \bq (x_{1 j}) \bq^\top (x_{1 j}) \big \{ \rho_{n,1} h_n (x_{1 j}, \btheta^*) - \rho_{n,1}^2 h_n^2 (x_{1 j}, \btheta^*) \big \} \label{hessian_dec2}. 
\end{align} 
We again study them term by term. 
For the term \eqref{hessian_dec1}, we can write 
\begin{align*} 
& n_1^{-1} \sum_{j = 1}^{n_0} \bq (x_{0 j}) \bq^\top (x_{0 j}) \big \{ \rho_{n,1} h_n (x_{0 j}, \btheta^*) - \rho_{n,1}^2 h_n^2 (x_{0 j}, \btheta^*) \big \} \\ 
= & \rho_{n,0} \Big \{ n_0^{-1} \sum_{j = 1}^{n_0} \bq (x_{0 j}) \bq^\top (x_{0 j}) h_n (x_{0 j}, \btheta^*) \Big \} 
- \rho_{n,0} \Big \{ n_0^{-1} \sum_{j = 1}^{n_0} \bq (x_{0 j}) \bq^\top (x_{0 j}) \rho_{n,1} h_n^2 (x_{0 j}, \btheta^*) \Big \}. 
\end{align*} 

The term in the first curly brackets is $ \bbE_0 [\bq (X) \bq^{\top} (X) h_n (X, \btheta^*)] + o_p (1) $ by the weak law of large numbers for triangular arrays \citep[Theorem 2.2.6]{durrett2019probability}, 
which further is $ \bbE_1 [\bq (X) \bq^{\top} (X)] + o_p (1) $ by the dominated convergence theorem as $ h_n (x, \btheta^*) \to \exp \{ \btheta^{* \top} \bq (x) \} $. 
Here we provide more details because the similar derivations will be seen many times in the future proof. 
Note that $ h_n (x, \btheta^*) < \rho_{n,0}^{-1} \exp \{ \btheta^{* \top} \bq (x) \} $, and $ \rho_{n,0}^{-1} \to 1 $ is bounded by some constant $ C $ uniformly for all large enough $ n $. 
Therefore, $ \| \bq (x) \bq^\top (x) h_n (x, \btheta^*) \| $ is uniformly bounded by $ C \| \bq (x) \|^2 \exp \{ \btheta^{* \top} \bq (x) \} $, which has finite expectation by Condition~\eqref{Condition.iii}.  

Similarly, the term in the second curly brackets is $ \bbE_0 [\bq (X) \bq^{\top} (X) \rho_{n,1} h_n^2 (X, \btheta^*)] + o_p (1) $ by the weak law of large numbers for triangular arrays, 
which further is $ o_p (1) $ by the dominated convergence theorem because $ \rho_{n,1} h_n^2 (x, \btheta^*) $ is uniformly bounded by $ C \exp \{ \btheta^{* \top} \bq (x) \} $ and converges pointwise to $ 0 $.
{\it (Remark: we can not factor out $ \rho_{n,1} $ and apply the law of large numbers to the remaining part here because the associated variance may not be finite.)} 

Combining these results leads to that \eqref{hessian_dec1} is $ \bbE_1 [\bq (X) \bq^{\top} (X)] + o_p (1) $. 
For the term \eqref{hessian_dec2}, because $ \rho_{n,1} h_n (x_{1 j}, \btheta^*) \to 0 $ and is uniformly bounded by $ 1 $, we get \eqref{hessian_dec2} is $ o_p (1) $. 
Therefore, we finally have 
\[ 
-n_1^{-1} \frac {\partial^2 \ell_{n} (\btheta^{*})} {\partial \btheta \partial \btheta^{\top}} 
= \bbE_1 [\bq (X) \bq^{\top} (X)] + o_p (1), 
\] 
which is positive definite. 
This proves the third conclusion in Lemma~\ref{lem:profile}. 

At this stage, we have completed the proof of Lemma~\ref{lem:profile}. 

\end{proof}

\subsection{Proof of Theorem~\ref {thm:para_normality}} 

This theorem states that the MELE $ \hat \btheta $, which is the maximizer of the profile log-EL $ \ell_{n} (\btheta) $, is asymptotically normal. 

\begin{proof}

We first give a rough order assessment on $ \hat \btheta - \btheta^* $ by showing that $ \hat \btheta - \btheta^* = O_p (n_1^{-1/3}) $, 
whose order will be refined later. 
The plan is as follows. 

Since $ \ell_{n} (\btheta) $ is a smooth function, there must be a maximizer of $ \ell_{n} (\btheta) $ in the compact set $ \{ \btheta: \| \btheta - \btheta^{*} \| \leq n_1^{-1/3} \} $. 
We prove that this maximizer is attained in the interior of the compact set with high probability. 
This will be done by showing that with high probability,
$ \ell_{n} (\btheta) < \ell_{n} (\btheta^{*}) $ uniformly for $ \btheta $ on the boundary of the compact set. 
In other words, this maximizer is a stationary point of the profile log-EL $ \ell_{n} (\btheta) $. 
Combined with the fact that the profile log-EL $ \ell_{n} (\btheta) $ is a concave function, 
this maximizer must coincide with the global maximizer of $ \ell_{n} (\btheta) $, which is $ \hat \btheta $. 
This conclusion would lead to  $ \hat \btheta - \btheta^* = O_p (n_1^{-1/3}) $. 

We now proceed to prove $ \hat \btheta - \btheta^* = O_p (n_1^{-1/3}) $. 
For any unit vector $ \ba $ and $ \btheta = \btheta^{*} + n_1^{-1/3} \ba $, expanding $ \ell_{n} (\btheta) $ at $ \btheta^{*} $ yields 
\begin{align} 
\ell_{n} (\btheta) = \ell_{n} (\btheta^{*}) 
+ n_1^{-1/3} \frac {\partial \ell_{n} (\btheta^{*})} {\partial \btheta} \ba 
+ \frac {1} {2} n_1^{-2/3} \ba^{\top} \frac {\partial^{2} \ell_{n} (\btheta^{*})} {\partial \btheta \partial \btheta^{\top}} \ba + \varepsilon_{n},  
\label {L_Expansion_1} 
\end{align} 
where $ \varepsilon_{n} $ is the remainder term: 
\begin{align*} 
\varepsilon_{n} = n_1^{-1} \sum_{ |\alpha| = 3 } \frac {1} {\alpha!} \frac {\partial^{\alpha} \ell_{n} (\underline {\btheta})} {\partial \btheta^{\alpha}} \, \, \ba^{\alpha}, 
\end{align*} 
for some $ \underline {\btheta} $ between $ \btheta^{*} $ and $ \btheta $. 
Note that $\alpha$ with $ |\alpha| = 3 $ is a vector of nonnegative integers with entries summing to 3.
Examples include $\alpha = (1, 0, 1, 0, 1)^\top$ and $\alpha = (2, 0, 1, 0, 0)^\top$ when $d = 5$. 
The partial derivative $ \partial^\alpha \ell_{n} (\btheta)/ \partial \btheta^\alpha $ is then with respect to $\theta_j$ with the corresponding order.

We first show that the remainder term $ \varepsilon_{n} $ in \eqref {L_Expansion_1} is $  O_p (1) $ uniformly over $ \ba $. 
We achieve this goal by showing that 
\begin{align*} 
\frac {\partial^{\alpha} \ell_{n} (\underline {\btheta})} {\partial \btheta^{\alpha}} = O_p (n_1), 
\end{align*} 
 for $ |\alpha| = 3 $ and uniformly over $ \ba $. 

Note that the partial derivative is given by 
\begin{align*}
\frac {\partial^{\alpha} \ell_{n} (\underline {\btheta})} {\partial \btheta^{\alpha}} 
= - \sum_{k, j} \frac {\rho_{n, 0} \rho_{n, 1} \exp \{ \underline {\btheta}^{\top} \bq (x_{k j}) \} [\rho_{n, 0} - \rho_{n, 1} \exp \{ \underline {\btheta}^{\top} \bq (x_{k j}) \}]} {[\rho_{n, 0} + \rho_{n, 1} \exp \{ \underline {\btheta}^{\top} \bq (x_{k j}) \}]^3} \bq^{\alpha} (x_{k j}), 
\end{align*} 
which is a sum of two terms indexed by $ k = 0, 1 $. 
The term with $ k = 1 $ is $ O_p (n_1) $ by the law of large numbers and the fact that 
\[ 
\left | 
\frac {\rho_{n, 0} \rho_{n, 1} \exp \{  \underline {\btheta}^{\top} \bq (x_{k j}) \} [\rho_{n, 0} - \rho_{n, 1} \exp \{  \underline {\btheta}^{\top} \bq (x_{k j}) \}]} {[\rho_{n, 0} + \rho_{n, 1} \exp \{  \underline {\btheta}^{\top} \bq (x_{k j}) \}]^3} 
\right |
\leq 1. 
\] 
Further, because 
\[ 
\left | 
\frac {\rho_{n, 0} \rho_{n, 1} \exp \{  \underline {\btheta}^{\top} \bq (x_{k j}) \} [\rho_{n, 0} - \rho_{n, 1} \exp \{  \underline {\btheta}^{\top} \bq (x_{k j}) \}]} {[\rho_{n, 0} + \rho_{n, 1} \exp \{  \underline {\btheta}^{\top} \bq (x_{k j}) \}]^3} 
\right |
\leq \rho_{n, 0}^{-1} \rho_{n, 1} \exp \{  \underline {\btheta}^{\top} \bq (x_{k j}) \},  
\] 
the term with $ k = 0 $ is 
\begin{align}
& \left | - \sum_{j=1}^{n_0} \frac {\rho_{n, 0} \rho_{n, 1} \exp \{ \underline {\btheta}^{\top} \bq (x_{0 j}) \} [\rho_{n, 0} - \rho_{n, 1} \exp \{ \underline {\btheta}^{\top} \bq (x_{0 j}) \}]} {[\rho_{n, 0} + \rho_{n, 1} \exp \{ \underline {\btheta}^{\top} \bq (x_{0 j}) \}]^3} \bq^{\alpha} (x_{0 j}) \right | \nonumber \\ 
= & O (\rho_{n, 0}^{-1} \rho_{n, 1}) \sum_{j=1}^{n_0} \exp \{ \underline {\btheta}^{\top} \bq (x_{0 j}) \} | \bq^{\alpha} (x_{0 j}) | \nonumber \\ 
= & O (n_1) n_0^{-1} \sum_{j=1}^{n_0} \exp \{ \underline {\btheta}^{\top} \bq (x_{0 j}) \} | \bq^{\alpha} (x_{0 j}) | \label {L_Expansion_1_rmd2} \\ 
= & O_p (n_1) \nonumber.
\end{align} 
The last equality is true because by the law of large numbers for triangular arrays and Condition~\eqref{Condition.iii} (which implies the associated variance is finite), we have 
\[ 
n_0^{-1} \sum_{j=1}^{n_0} \exp \{ \underline {\btheta}^{\top} \bq (x_{0 j}) \} | \bq^{\alpha} (x_{0 j}) | 
= \bbE_{0} \left [ \exp \{ \underline {\btheta}^{\top} \bq (X) \} | \bq^{\alpha} (X) | \right ] + o_p (1). 
\] 
We remark that without loss of generality, we treat $ \underline{\btheta} $ as if it is sample-independent that is uniformly $ \| \underline {\btheta} - \btheta^* \| \leq O_p (n_1^{-1/3}) $. 
Further, by Condition~\eqref{Condition.iii} again and the Cauchy--Schwarz inequality, the main term in the right hand side of the preceding equality is uniformly bounded for $ \underline {\btheta} $ in a neighbourhood of $ \btheta^{*} $. 
Therefore, the main term in \eqref {L_Expansion_1_rmd2} is $ O_p (1) $. 

Thus, we have shown that
\begin{align}
\frac {\partial^{\alpha} \ell_{n} (\underline {\btheta})} {\partial \btheta^{\alpha}} = O_p (n_1) 
\label{L_Expansion_1_rmd3}
\end{align}
for $ | \alpha | = 3 $, and $ \varepsilon_{n} = O_p (1) $ uniformly over $ \ba $. 

Next, we proceed with the main terms in the expansion in \eqref {L_Expansion_1}.
For the first derivative term in \eqref {L_Expansion_1}, the asymptotic normality conclusion in Lemma~\ref {lem:profile} implies 
\begin{align*}
\frac {\partial \ell_{n} (\btheta^{*})} {\partial \btheta}  
= O_p (n_1^{1/2}) 
= o_p (n_1^{2/3}).
\end{align*} 
For the second derivative term in \eqref {L_Expansion_1}, the Hessian conclusion in Lemma~\ref {lem:profile} implies 
\begin{align*} 
\frac {\partial^{2} \ell_{n} (\btheta^{*})} {\partial \btheta \partial \btheta^{\top}} 
= -n_1 \left \{ \bbE_{1} [ \bq \bq^{\top} ] + o_p (1) \right \}. 
\end{align*}
Combining these order assessments for terms in \eqref {L_Expansion_1}, we find
\begin{align*} 
\ell_{n} (\btheta^{*} + n_1^{-1/3} \ba) - \ell_{n} (\btheta^{*}) 
& = n_1^{1/3} \ba^{\top} \left \{ -\bbE_{1} [ \bq \bq^{\top} ] + o_p (1) \right \} \ba + o_p (n_1^{1/3}) \\
& = n_1^{1/3} \{ -C + o_p (1) \}, 
\end{align*} 
for some positive constant $ C $ because $ \bbE_{1} [ \bq \bq^{\top} ] $ is positive definite. 

Note that the event $ \| \hat \btheta - \btheta^{*} \| < n_1^{-1/3} $ 
is implied by the event $ \ell_{n} (\btheta^{*} + n_1^{-1/3} \ba) < \ell_{n} (\btheta^{*}) $. 
Therefore, with the same positive constant $ C $, 
we have 
\begin{align*} 
P ( \| \hat \btheta - \btheta^{*} \| < n_1^{-1/3} ) 
& \geq 
P (\ell_{n} (\btheta^{*} + n_1^{-1/3} \ba) < \ell_{n} (\btheta^{*})) \\ 
& = 
P (o_p (1) < C), 
\end{align*} 
which is arbitrarily close to 1. 
This proves that the unique maximizer $ \hat \btheta $ of $ \ell_{n} (\btheta) $ satisfies  
\begin{align} 
\hat \btheta - \btheta^{*} = O_p (n_1^{-1/3}). 
\label {thetahat_order} 
\end{align} 

We now prove the asymptotic normality of $ \hat \btheta $. 
Expanding $ \partial \ell_{n} (\hat \btheta) / \partial \btheta $ at the truth $ \btheta^{*} $, we get  
\begin{align} 
\mathbf {0} = \frac {\partial \ell_{n} (\hat \btheta)} {\partial \btheta} 
& = \frac {\partial \ell_{n} (\btheta^{*})} {\partial \btheta} 
+ \frac {\partial^{2} \ell_{n} (\btheta^{*})} {\partial \btheta \partial \btheta^{\top}} 
(\hat \btheta - \btheta^{*}) 
+ \varepsilon'_{n},  
\label {Taylor_E3}
\end{align}  
where $ \varepsilon'_{n} $ is the remainder term: 
\begin{align*} 
\varepsilon'_{n} 
= \sum_{ |\alpha| = 2 } \frac {1} {\alpha!} \frac {\partial^{\alpha+1} \ell_{n} (\bar \btheta)} {\partial \btheta^{\alpha+1}} 
(\hat \btheta - \btheta^{*})^{\alpha}, 
\end{align*} 
for some $ \bar \btheta $ between $ \btheta^{*} $ and $ \hat \btheta $. 
By the same technique as we earlier proved \eqref{L_Expansion_1_rmd3}, 
we have 
\[ 
\frac {\partial^{\alpha+1} \ell_{n} (\bar \btheta)} {\partial \btheta^{\alpha+1}}
= O_p (n_1) 
\] 
for $ |\alpha| = 2 $. 
Further, because $ \hat \btheta - \btheta^{*} = O_p (n_1^{-1/3}) $ 
as shown in \eqref {thetahat_order} , 
we have 
\[ 
(\hat \btheta - \btheta^{*})^{\alpha}
= 
O_p (n_1^{-2/3}).
\]  
These results lead to $ \varepsilon'_{n} = O_p (n_1^{1/3}) $. 

We have shown in Lemma~\ref {lem:profile} that 
\[ 
- n_1^{-1} \frac {\partial^{2} \ell_{n} (\btheta^*)} {\partial \btheta \partial \btheta^{\top}} 
= 
\bbE_{1} [ \bq \bq^{\top} ] + o_p (1). 
\] 
Therefore, rearranging the equation in \eqref {Taylor_E3} yields
\begin{align} 
\left [ \bbE_{1} [ \bq \bq^{\top} ] + o_p (1) \right ] \sqrt {n_1} 
(\hat \btheta - \btheta^{*}) 
= n_1^{-1/2} \frac {\partial \ell_{n} (\btheta^{*})} {\partial \btheta} + o_p (1). 
\label {Taylor_E3_rearrange}
\end{align} 
By the asymptotic normality conclusion in Lemma~\ref {lem:profile}, we have 
\[
\frac {\partial \ell_{n} (\btheta^*)} {\partial \btheta} 
= 
O_p (n_1^{1/2}). 
\]
Therefore, it must be true that $ \hat \btheta - \btheta^{*} = O_{p} (n_1^{-1/2}) $; 
otherwise, the orders in the two sides of the equation \eqref {Taylor_E3_rearrange} would not match. 
With this refinement of the order of $ \hat \btheta - \btheta^{*} $, 
we have 
\begin{align*} 
\sqrt {n_1} (\hat \btheta - \btheta^{*})
& = \{ \bbE_{1} [ \bq \bq^{\top} ] \}^{-1} 
\left [ n_1^{-1/2} 
\frac {\partial \ell_{n} (\btheta^{*})} {\partial \btheta} 
\right ] 
+ o_p (1). 
\end{align*} 

Hence, by Slutsky's Theorem, we finally get
\begin{align} 
\sqrt {n_1} (\hat \btheta - \btheta^{*})
& \overset {d} \to 
N \left (\mathbf {0}, \{ \bbE_{1} [ \bq \bq^{\top} ] \}^{-1} \Var_1 [\bq] \{ \bbE_{1} [ \bq \bq^{\top} ] \}^{-1} \right), 
\label {thm:para_normality_theta}
\end{align} 
as $ n_0, n_1 \to \infty $, 
where we recall that $ \Var_1 [\bq] $ is the covariance matrix in the limiting normal distribution of the score function
$ n_1^{-1/2} \partial \ell_{n} (\btheta^*)/\partial \btheta $.

We now simplify the covariance matrix in \eqref {thm:para_normality_theta}. 
Because $ \bq^{\top} (x) = (1, \bq_{-}^{\top} (x)) $, 
we have 
\[ 
\Var_1 [\bq (X)] = 
\begin{pmatrix}
0 & \mathbf {0} \\ 
\mathbf {0} & \Var_1 [\bq_{-} (X)] 
\end{pmatrix}.  
\] 
Also, 
\[ 
\mathbbm {E}_{1} [\bq (X) \bq^{\top} (X)] 
= 
\begin{pmatrix}
1 & \mathbbm {E}_{1} [\bq_{-}^{\top} (X)] \\ 
\mathbbm {E}_{1} [\bq_{-} (X)] & \mathbbm {E}_{1} [\bq_{-} (X) \bq_{-}^{\top} (X)] 
\end{pmatrix}, 
\] 
whose inverse is given by (see Theorem 8.5.11 in \citet {harville1997matrix}) 
\begin{align*}
\{ \mathbbm {E}_{1} [\bq (X) \bq^{\top} (X)] \}^{-1} 
= 
\begin{pmatrix}
1 & \mathbf {0} \\ 
\mathbf {0} & \mathbf {0} 
\end{pmatrix} 
+
\begin{pmatrix}
- \bbE_1 [\bq_{-}^{\top} (X)] \\ 
\bI_d
\end{pmatrix}
\Var_1^{-1} [\bq_{-} (X)]
\begin{pmatrix}
- \bbE_1 [\bq_{-} (X)], 
\, 
\bI_d
\end{pmatrix}. 
\end{align*}
These matrix results lead to a more friendly expression of the covariance matrix in \eqref {thm:para_normality_theta}: 
\begin{align*} 
& \{ \mathbbm {E}_{1} [\bq (X) \bq^{\top} (X)] \}^{-1} \Var_1 [\bq (X)] \{ \mathbbm {E}_{1} [\bq (X) \bq^{\top} (X)] \}^{-1} \\ 
& = 
\begin{pmatrix}
- \bbE_1 [\bq_{-}^{\top} (X)] \\ 
\bI_d
\end{pmatrix}
\Var_1^{-1} [\bq_{-} (X)]
\begin{pmatrix}
- \bbE_1 [\bq_{-} (X)], 
\, 
\bI_d
\end{pmatrix} \\ 
& = 
\{ \mathbbm {E}_{1} [\bq (X) \bq^{\top} (X)] \}^{-1} - 
\begin{pmatrix}
1 & \mathbf {0} \\ 
\mathbf {0} & \mathbf {0} 
\end{pmatrix}.
\end{align*} 

This completes the proof of Theorem~\ref {thm:para_normality}. 

\end{proof}

\subsection{Proof of Theorem~\ref{thm:dist_normality}} 

This theorem asserts that for every $ x $ in the support of the distribution function $ G_1 $, 
the corresponding DRM-based distribution estimator $ \hat G_1 (x) $ is asymptotically normal. 

\begin{proof}

Following \citet[proof of Theorem 3.2] {chen2013quantile}, we can write $ \hat G_1 (x) $ and $ G_1 (x) $ as 
\begin{align*} 
\hat G_1 (x) 
= & n_1^{-1} \sum_{k = 0, 1} \sum_{j = 1}^{n_k} 
\frac {\rho_{n, 1} \exp \{ \hat \btheta^{\top} \bq (x_{k j}) \}} {\rho_{n, 0} + \rho_{n, 1} \exp \{ \hat \btheta^{\top} \bq (x_{k j}) \}} \ind (x_{k j} \leq x), 
\\ 
G_1 (x) 
= & n_1^{-1} \sum_{k = 0, 1} \sum_{j = 1}^{n_k} \bbE_k 
\left [
\frac {\rho_{n, 1} \exp \{ \btheta^{* \top} \bq (x_{k j}) \}} {\rho_{n, 0} + \rho_{n, 1} \exp \{ \btheta^{* \top} \bq (x_{k j}) \}} \ind (x_{k j} \leq x)
\right ], 
\end{align*} 
where for each $ k $, the variables $ \{ x_{k j}: j =1, \ldots, n_k \} $ are taken expectation with respect to $ G_k $. 
Recall that we defined earlier: 
\[ 
h_n (x, \btheta) = \frac {\exp \{ \btheta^{\top} \bq (x) \}} {\rho_{n, 0} + \rho_{n, 1} \exp \{ \btheta^{\top} \bq (x) \}}. 
\] 
Therefore, we may write their difference as 
\begin{align} 
\hat G_1 (x) - G_1 (x)  
= & n_1^{-1} \sum_{k, j} \rho_{n, 1}
\Big \{ 
h_n (x_{k j}, \hat \btheta) \ind (x_{k j} \leq x) - \bbE_k [h_n (x_{k j}, \btheta^*) \ind (x_{k j} \leq x)]
\Big \} 
\nonumber \\ 
= & n_1^{-1} \sum_{k, j} \rho_{n, 1} 
\Big \{ 
h_n (x_{k j}, \btheta^*) \ind (x_{k j} \leq x) - \bbE_k [h_n (x_{k j}, \btheta^*) \ind (x_{k j} \leq x)]
\Big \} 
\label {thm:dist_normality_proof1}
\\ 
+ & (\hat \btheta - \btheta^{*})^{\top} 
\Big \{ n_1^{-1} \sum_{k, j} \frac {\rho_{n, 0} \rho_{n, 1} h_n (x_{k j}, \btheta^*) \bq (x_{k j})} {\rho_{n, 0} + \rho_{n, 1} \exp \{ \btheta^{* \top} \bq (x_{k j}) \}} \ind (x_{k j} \leq x) 
\Big \} 
+ R_n. 
\label {thm:dist_normality_proof2}
\end{align} 
The remainder term is $ R_n = o_p (n_1^{-1/2}) $, but we postpone this claim to the end of this section. 

Notice that \eqref{thm:dist_normality_proof1} is a sum of two terms indexed by $ k = 0, 1 $, which are respectively similar to \eqref{score_dec1} and \eqref{score_dec2} that we handled earlier in the proof of Lemma~\ref{lem:profile}. 
Therefore, we omit some details here and conclude directly based on the central limit theorem for triangular arrays and the dominated convergence theorem that \eqref {thm:dist_normality_proof1} is $ o_p (n_1^{-1/2}) $. 

Next, we deal with the fraction term in \eqref {thm:dist_normality_proof2}, which is also a sum of two terms indexed by $ k = 0, 1 $. 
The term with $ k = 0 $ is 
\begin{align*} 
& n_1^{-1} \sum_{j=1}^{n_0} \frac {\rho_{n, 0} \rho_{n, 1} h_n (x_{0 j}, \btheta^*) \bq (x_{0 j})} {\rho_{n, 0} + \rho_{n, 1} \exp \{ \btheta^{* \top} \bq (x_{0 j}) \}} \ind (x_{0 j} \leq x) \\ 
= & \rho_{n, 0}^2 n_0^{-1} \sum_{j=1}^{n_0} \frac {h_n (x_{0 j}, \btheta^*) \bq (x_{0 j})} {\rho_{n, 0} + \rho_{n, 1} \exp \{ \btheta^{* \top} \bq (x_{0 j}) \}} \ind (x_{0 j} \leq x) \\ 
= & \bbE_0 \left [ \frac {\rho_{n, 0}^2 h_n (X, \btheta^*) \bq (X)} {\rho_{n, 0} + \rho_{n, 1} \exp \{ \btheta^{* \top} \bq (X) \}} \ind (X \leq x) \right ] + o_p (1) \\ 
= & \bbE_0 [\exp \{ \btheta^{* \top} \bq (X) \} \bq (X) \ind (X \leq x)] + o_p (1) \\ 
= & 
\begin{pmatrix}
G_1 (x) \\ 
\bQ (x)
\end{pmatrix}
+ o_p (1).
\end{align*} 
Note that the third last equality is by applying the weak law of large numbers for triangular arrays; 
the second last equality is by the dominated convergence theorem because the term within the expectation is uniformly bounded by $ \| \bq (X) \| \exp \{ \btheta^{* \top} \bq (X) \} $ that has finite expectation, and $ \rho_{n,0} \to 1, \rho_{n,1} \to 0 $;   
and the last equality is by decomposing $ \bq^\top (x) = (1, \bq^\top (x)) $, with $ \bQ (x) $ defined in the main Theorem~\ref {thm:dist_normality}. 
We can apply the same technique for the term with $ k = 1 $: 
\begin{align*} 
& n_1^{-1} \sum_{j=1}^{n_1} \frac {\rho_{n, 0} \rho_{n, 1} h_n (x_{1 j}, \btheta^*) \bq (x_{1 j})} {\rho_{n, 0} + \rho_{n, 1} \exp \{ \btheta^{* \top} \bq (x_{1 j}) \}} \ind (x_{1 j} \leq x) \\ 
= & \bbE_1 \left [ \frac {\rho_{n, 0} \rho_{n, 1} h_n (X, \btheta^*) \bq (X)} {\rho_{n, 0} + \rho_{n, 1} \exp \{ \btheta^{* \top} \bq (X) \}} \ind (X \leq x) \right ] + o_p (1) 
= o_p (1).
\end{align*} 
Therefore, we have shown that the fraction term in \eqref {thm:dist_normality_proof2} is $ (G_1(x), \, \bQ^{\top} (x))^\top + o_p (1) $. 

Combining these results for \eqref {thm:dist_normality_proof1}--\eqref {thm:dist_normality_proof2} and $ \hat \btheta - \btheta^{*} = O_p (n_1^{-1/2}) $ from Theorem~\ref{thm:para_normality} yields  
\[ 
\hat G_1 (x) - G_1 (x) = 
(G_1(x), \, \bQ^{\top} (x)) \{ \hat \btheta - \btheta^{*} \} + o_p (n_1^{-1/2}).  
\] 
Because we have proved in Theorem~\ref{thm:para_normality} that $ n_1^{1/2} (\hat \btheta - \btheta^{*}) $ is asymptotically normal with covariance matrix in \eqref{thm:para_normality_mat},
it is clear that $ n_1^{1/2} \{ \hat G_1 (x) - G_1 (x) \} $ is also asymptotically normal with mean zero and variance  
\begin{align*}
& 
(G_1(x), \, \bQ^{\top} (x)) 
\begin{pmatrix}
- \bbE_1 [\bq_{-}^{\top}] \\ 
\bI_d
\end{pmatrix}
\Var_1^{-1} [\bq_{-}]
\begin{pmatrix}
- \bbE_1 [\bq_{-}], 
\, 
\bI_d
\end{pmatrix} 
\begin{pmatrix}
G_1 (x) \\ 
\bQ (x)
\end{pmatrix}
\\ 
& = 
\{ \bQ (x) - \bbE_1 [\bq_{-}] G_1 (x) \}^{\top}
\Var_1^{-1} [\bq_{-}]
\{ \bQ (x) - \bbE_1 [\bq_{-}] G_1 (x) \}.
\end{align*} 

So far, the main conclusion of the Theorem~\ref {thm:dist_normality} has been proved, except for the proof that the remainder term $ R_n $ in \eqref {thm:dist_normality_proof2} is $ o_p (n_1^{-1/2}) $.
The following is devoted to this task. 

For simplicity, we define 
\[ 
B_n (\btheta) = n_1^{-1} \sum_{k, j} \rho_{n, 1} h_n (x_{k j}, \btheta) \ind (x_{k j} \leq x). 
\]

For some $ \btheta_{\dagger} $ between $ \hat \btheta $ and $ \btheta^{*} $, we have 
\begin{align}
R_n = \frac {1} {2} (\hat \btheta - \btheta^{*})^{\top} 
\frac {\partial^2 B_n (\btheta^{*})} {\partial \btheta \partial \btheta^{\top}}
(\hat \btheta - \btheta^{*})
+ 
\sum_{ |\alpha| = 3 } \frac {1} {\alpha!} \frac {\partial^{\alpha} B_n (\btheta_{\dagger})} {\partial \btheta^{\alpha}} 
(\hat \btheta - \btheta^{*})^{\alpha}, 
\label {taylor_rmd_cdf}
\end{align} 
where 
\begin{align*}
\frac {\partial^2 B_n (\btheta^{*})} {\partial \btheta \partial \btheta^{\top}} 
& = n_1^{-1} \sum_{k, j} \frac {\rho_{n, 0} \rho_{n, 1} h_n (x_{k j}, \btheta^*) [\rho_{n, 0} - \rho_{n, 1} \exp \{ \btheta^{* \top} \bq (x_{k j}) \}]} {[\rho_{n, 0} + \rho_{n, 1} \exp \{ \btheta^{* \top} \bq (x_{k j}) \}]^2} \bq (x_{k j}) \bq^{\top} (x_{k j}) \ind (x_{k j} \leq x), 
\end{align*} 
and for $ |\alpha| = 3 $, 
\begin{align*}
\frac {\partial^{\alpha} B_n (\btheta_{\dagger})} {\partial \btheta^{\alpha}} 
& = n_1^{-1} \sum_{k, j} \frac {\rho_{n, 0} \rho_{n, 1} h_n (x_{k j}, \btheta_{\dagger}) [\rho_{n, 0} - 2 \rho_{n, 1} \exp \{ \btheta_{\dagger}^{\top} \bq (x_{k j}) \}]^2} {[\rho_{n, 0} + \rho_{n, 1} \exp \{ \btheta_{\dagger}^{\top} \bq (x_{k j}) \}]^3} \bq^{\alpha} (x_{k j}) \ind (x_{k j} \leq x). 
\end{align*} 
By the law of large numbers for triangular arrays, we have $ \partial^2 B_n (\btheta^{*})/\partial \btheta \partial \btheta^{\top} = O_p (1) $. 
Next, we show that $ \partial^{\alpha} B_n (\btheta_{\dagger})/\partial \btheta^{\alpha} = O_p (1) $ for $ |\alpha| = 3 $. 

Note that $ \partial^{\alpha} B_n (\btheta_{\dagger})/\partial \btheta^{\alpha} $ is a sum of two terms indexed by $ k = 0, 1 $. 
The term with $ k = 1 $ is $ O_p (1) $ by the law of large numbers and the fact that 
\[ 
\frac {\rho_{n, 0} \rho_{n, 1} h_n (x_{k j}, \btheta_{\dagger}) [\rho_{n, 0} - 2 \rho_{n, 1} \exp \{ \btheta_{\dagger}^{\top} \bq (x_{k j}) \}]^2} {[\rho_{n, 0} + \rho_{n, 1} \exp \{ \btheta_{\dagger}^{\top} \bq (x_{k j}) \}]^3}
\leq 4. 
\] 
Further, because 
\[ 
\frac {\rho_{n, 0} \rho_{n, 1} h_n (x_{k j}, \btheta_{\dagger}) [\rho_{n, 0} - 2 \rho_{n, 1} \exp \{ \btheta_{\dagger}^{\top} \bq (x_{k j}) \}]^2} {[\rho_{n, 0} + \rho_{n, 1} \exp \{ \btheta_{\dagger}^{\top} \bq (x_{k j}) \}]^3}
\leq 
4 \rho_{n, 0}^{-1} \rho_{n, 1} \exp \{ \btheta_{\dagger}^{\top} \bq (x_{0 j}) \},  
\] 
the term with $ k = 0 $ is 
\begin{align}
& \left | n_1^{-1} \sum_{j=1}^{n_0} 
\frac {\rho_{n, 0} \rho_{n, 1} h_n (x_{0 j}, \btheta_{\dagger}) [\rho_{n, 0} - 2 \rho_{n, 1} \exp \{ \btheta_{\dagger}^{\top} \bq (x_{0 j}) \}]^2} {[\rho_{n, 0} + \rho_{n, 1} \exp \{ \btheta_{\dagger}^{\top} \bq (x_{0 j}) \}]^3} \bq^{\alpha} (x_{0 j}) \ind (x_{0 j} \leq x) \right | \nonumber \\ 
= & O (\rho_{n, 0}^{-1} \rho_{n, 1}) n_1^{-1} \sum_{j=1}^{n_0} \exp \{ \btheta_{\dagger}^{\top} \bq (x_{0 j}) \} | \bq^{\alpha} (x_{0 j}) | \ind (x_{0 j} \leq x) \nonumber \\ 
= & O (1) n_0^{-1} \sum_{j=1}^{n_0} \exp \{ \btheta_{\dagger}^{\top} \bq (x_{0 j}) \} | \bq^{\alpha} (x_{0 j}) | \ind (x_{0 j} \leq x) \label {taylor_rmd_cdf2} \\ 
= & O_p (1) \nonumber.
\end{align} 
The last equality is true because by the law of large numbers for triangular arrays and Condition~\eqref{Condition.iii} (for finite variance), 
\[ 
n_0^{-1} \sum_{j=1}^{n_0} \exp \{ \btheta_{\dagger}^{\top} \bq (x_{0 j}) \} | \bq^{\alpha} (x_{0 j}) | \ind (x_{0 j} \leq x) 
= \bbE_{0} \left [ \exp \{ \btheta_{\dagger}^{\top} \bq (X) \} | \bq^{\alpha} (X) | \ind (X \leq x) \right ] + o_p (1). 
\] 
We note that without loss of generality, we regard $ \btheta_{\dagger} $ as sample-independent that is uniformly $ \| \btheta_{\dagger} - \btheta^* \| \leq O_p (n_1^{-1/2}) $. 
Further, by Condition~\eqref{Condition.iii} again and the Cauchy--Schwarz inequality, the main term in the right hand side of the preceding equality is uniformly bounded for $ \btheta_{\dagger} $ in a neighbourhood of $ \btheta^{*} $. 
Therefore, the term in \eqref {taylor_rmd_cdf2} is $ O_p (1) $. 

By plugging these rate results in \eqref {taylor_rmd_cdf}, along with that $ \hat \btheta - \btheta^{*} = O_p (n_1^{-1/2}) $, 
we have thus successfully proved that 
\[ 
R_n 
= O_p (n_1^{-1}) + O_p (n_1^{-3/2})
= o_p (n_1^{-1/2}). 
\] 

This completes the proof. 

\end{proof}

\subsection{Proof of Theorem~\ref{thm:quan_normality}} 

This theorem states that the DRM-based quantile estimator $ \hat \xi_p $ of the $ p $th quantile of distribution $ G_1 $ has a limiting normal distribution.

\begin{proof}

We prove the desired result by definition. 
First, note that 
\begin{align*} 
P ( \sqrt{n_1} \{ \hat \xi_p - \xi_p \} \leq t) 
= P (\hat \xi_p \geq \xi_p + t n_1^{-1/2})
= P \Big (\hat G_1(\xi_p + t n_1^{-1/2}) \geq p \Big )
\end{align*} 
by the equivalence between the last two events in the probabilities, which is implied by the definition of the quantile estimator $ \hat \xi_p $ in \eqref{DRM-quan}. 
Then, we have
\begin{align} 
& P ( \sqrt{n_1} \{ \hat \xi_p - \xi_p \} \leq t)
= P \Big (\hat G_1(\xi_p + t n_1^{-1/2}) \geq p \Big ) \nonumber \\
= & P \Big (\hat G_1(\xi_p + t n_1^{-1/2}) - G_1 (\xi_p + t n_1^{-1/2}) \geq G_1 (\xi_p) - G_1 (\xi_p + t n_1^{-1/2}) \Big ) \nonumber \\
= & P \Big ([g_1 (\xi_p)]^{-1} \sqrt{n_1} \{\hat G_1 (\xi_p + t n_1^{-1/2}) - G_1 (\xi_p + t n_1^{-1/2}) \} \geq t + o (1) \Big ) \label{quan_proof1}, 
\end{align} 
because the density function $ g_1 (\cdot) $ is continuous and positive at $ \xi_p $. 

We now work on the distribution terms in \eqref{quan_proof1}. 
Following the same line of the proof of Theorem~\ref {thm:dist_normality}, with a generic $ x $ we write 
\begin{align} 
& \hat G_1 (x) - G_1 (x) \nonumber \\ 
= & n_1^{-1} \sum_{k, j} \rho_{n, 1} 
\Big \{ 
h_n (x_{k j}, \btheta^*) \ind (x_{k j} \leq x) - \bbE_k [h_n (x_{k j}, \btheta^*) \ind (x_{k j} \leq x)]
\Big \} 
\label{quan_proof2}
\\ 
+ & (\hat \btheta - \btheta^{*})^{\top} 
\Big \{ n_1^{-1} \sum_{k, j} \frac {\rho_{n, 0} \rho_{n, 1} h_n (x_{k j}, \btheta^*) \bq (x_{k j})} {\rho_{n, 0} + \rho_{n, 1} \exp \{ \btheta^{* \top} \bq (x_{k j}) \}} \ind (x_{k j} \leq x) 
\Big \} 
+ R_n, 
\label{quan_proof3}
\end{align} 
where 
\[ 
h_n (x, \btheta) = \frac {\exp \{ \btheta^{\top} \bq (x) \}} {\rho_{n, 0} + \rho_{n, 1} \exp \{ \btheta^{\top} \bq (x) \}}. 
\] 
When $ x $ is fixed (independent of $ n $), we have shown in the proof of Theorem~\ref {thm:dist_normality} that: 
\begin{enumerate} 
\item the term \eqref{quan_proof2} is $ o_p (n_1^{-1/2}) $; 
\item the fraction term in \eqref{quan_proof3} is $ (G_1(x), \, \bQ^{\top} (x))^\top + o_p (1) $; and 
\item the remainder term $ R_n $ is $ o_p (n_1^{-1/2}) $. 
\end{enumerate} 
When $ x = \xi_p + t n_1^{-1/2} \to \xi_p $, the same argument can be made for the above items 1 and 3, and for item 2: 
\[ 
n_1^{-1} \sum_{k, j} \frac {\rho_{n, 0} \rho_{n, 1} h_n (x_{k j}, \btheta^*) \bq (x_{k j})} {\rho_{n, 0} + \rho_{n, 1} \exp \{ \btheta^{* \top} \bq (x_{k j}) \}} \ind (x_{k j} \leq \xi_p + t n_1^{-1/2}) 
= 
\begin{pmatrix}
p \\ 
\bQ (\xi_p)
\end{pmatrix}
+ o_p (1).
\] 
These claims can all be proved by applying the same techniques as we used earlier in the proof of Theorem~\ref {thm:dist_normality} for handling \eqref{thm:dist_normality_proof1}--\eqref{thm:dist_normality_proof2}, and hence we omit some details here. 

These rate results, along with the conclusions from Theorem~\ref{thm:para_normality} and \eqref{thm:para_normality_mat} on the asymptotic normality of $ \sqrt{n_1} (\hat \btheta - \btheta^{*}) $, 
lead to that the main term in \eqref{quan_proof1} is also asymptotically normal: 
\begin{align}
[g_1 (\xi_p)]^{-1} \sqrt{n_1} & \{\hat G_1 (\xi_p + t n_1^{-1/2}) - G_1 (\xi_p + t n_1^{-1/2}) \} \nonumber \\ 
& \overset{d} \to N \left (0, 
\{ \bQ (\xi_p) - p \bbE_1 [\bq_{-}] \}^{\top}
\frac {\Var_1^{-1} [\bq_{-}]}{g_1^2 (\xi_p)}
\{ \bQ (\xi_p) - p \bbE_1 [\bq_{-}] \} \right ). 
\label{quan_proof4}
\end{align} 
Therefore, by Slutsky's theorem, the $ o(1) $ term in \eqref{quan_proof1} can be ignored, and the desired probability $ P (\sqrt{n_1} \{ \hat \xi_p - \xi_p \} \leq t) $ converges to the cumulative distribution function of the normal distribution in \eqref{quan_proof4}. 

This completes the proof. 

\end{proof}

\subsection{Proof of Theorem~\ref {thm:bahadur}} 

This theorem presents the Bahadur representation for the DRM-based quantile estimator $ \hat \xi_p $ at quantile level $ p $ for distribution $ G_1 $. 

\begin{proof}

We first claim a rate result on distribution estimator $ \hat G_1 $, as follows. 
\begin{align} 
\sup_{|x - \xi_p| \leq O_p (n_1^{-1/2})} 
| \{ \hat G_1 (x) - \hat G_1 (\xi_p) \} 
- \{ G_1 (x) - G_1 (\xi_p) \} | 
= O_p (n_1^{-3/4} \log^{1/2} n_1). 
\label {lem:bahadur_claim}
\end{align} 
For a better presentation, we delay the proof of the claim in \eqref {lem:bahadur_claim} to the end of this proof. 

We proceed to show the Bahadur representation. 

Because the density $ g_1 (\cdot) $ is continuous at $ \xi_p $, we have 
\[ 
G_1 (\hat \xi_p) - G_1 (\xi_p) 
= g_1 (\xi_p) \{ \hat \xi_p - \xi_p \} + O_p (n_1^{-1}).
\] 
Since $ \hat G_1 (x) $ is a discrete step function that has an increment of size $ O (n_1^{-1}) $ at each observation (jump point), by the definition of $ \hat \xi_p $, we have $ \hat G_1 (\hat \xi_p) - p = O_p (n_1^{-1}) $. 
Further, we have shown in Theorem~\ref{thm:quan_normality} that $ \hat \xi_p - \xi_p = O_p (n_1^{-1/2}) $. 
Combing these results and by letting $ x $ be $ \hat \xi_p $ in the claim in \eqref {lem:bahadur_claim}, we have 
\begin{align*} 
\{ p - \hat G_1 (\xi_p) \} - g_1 (\xi_p) \{ \hat \xi_p - \xi_p \} + O_p (n_1^{-1}) 
= O_p (n_1^{-3/4} \log^{1/2} n_1), 
\end{align*} 
which, after rearrangement, leads to the desired result: 
\begin{align*} 
\hat \xi_p - \xi_p 
= \frac {G_1 (\xi_p) - \hat G_1 (\xi_p)} {g_1 (\xi_p)} 
+ O_p (n_1^{-3/4} \log^{1/2} n_1). 
\end{align*} 

Finally, we now present the proof of the claim in \eqref {lem:bahadur_claim}. 
The proof is similar to the proof of Lemma A.2 in \citet {chen2013quantile}, with some modifications. 
Without loss of generality, we assume that $ x \geq \xi_p $. 

Recall that $ \hat G_1 (x) $ and $ G_1 (x) $ can be written as 
\begin{align*} 
\hat G_1 (x) 
= & n_1^{-1} \sum_{k = 0, 1} \sum_{j = 1}^{n_k} \rho_{n, 1} h_n (x_{k j}, \hat \btheta) \ind (x_{k j} \leq x), 
\\ 
G_1 (x) 
= & n_1^{-1} \sum_{k = 0, 1} \sum_{j = 1}^{n_k} \bbE_k [\rho_{n, 1} h_n (x_{k j}, \btheta^{*}) \ind (x_{k j} \leq x)], 
\end{align*} 
where 
\[ 
h_n (x, \btheta) = \frac {\exp \{ \btheta^{\top} \bq (x) \}} {\rho_{n, 0} + \rho_{n, 1} \exp \{ \btheta^{\top} \bq (x) \}}. 
\] 
We also define another function by replacing $ \btheta $ with $ \btheta^{*} $ in $ \hat G_1 $: 
\[ 
G_1^{*} (x) 
= n_1^{-1} \sum_{k = 0, 1} \sum_{j = 1}^{n_k} \rho_{n, 1} h_n (x_{k j}, \btheta^{*}) \ind (x_{k j} \leq x). 
\] 
We then rewrite the left-hand side of \eqref {lem:bahadur_claim} as 
\begin{align} 
& \{ \hat G_1 (x) - \hat G_1 (\xi_p) \} 
- \{ G_1 (x) - G_1 (\xi_p) \} 
\nonumber \\ 
= & \{ \hat G_1 (x) - \hat G_1 (\xi_p) \} 
- \{ G_1^{*} (x) - G_1^{*} (\xi_p) \} 
\label {lem:bahadur_claim_rewrite1} \\ 
& + 
\{ G_1^{*} (x) - G_1^{*} (\xi_p) \} 
- \{ G_1 (x) - G_1 (\xi_p) \} . 
\label {lem:bahadur_claim_rewrite2}
\end{align} 
Next, we deal with the terms \eqref {lem:bahadur_claim_rewrite1} and \eqref {lem:bahadur_claim_rewrite2} one by one. 

For the term \eqref {lem:bahadur_claim_rewrite2}, 
because $ \bbE [G_1^{*} (x)] = G_1 (x) $ and $ G_1^{*} $ is also a distribution function, 
following the proof in \citet [Lemma 2.5.4E] {serfling2000approximation}, 
we have 
\begin{align} 
\sup_{|x - \xi_p| \leq O_p (n_1^{-1/2})} 
| \{ G_1^{*} (x) - G_1^{*} (\xi_p) \} 
- \{ G_1 (x) - G_1 (\xi_p) \} |
= O_p (n_1^{-3/4} \log^{1/2} n_1). 
\label {lem:bahadur_claim_rewrite2_order}
\end{align} 

For the term \eqref {lem:bahadur_claim_rewrite1}, we note that 
\begin{align} 
& \{ \hat G_1 (x) - \hat G_1 (\xi_p) \} 
- \{ G_1^{*} (x) - G_1^{*} (\xi_p) \} 
\nonumber \\ 
= & n_1^{-1} \sum_{k, j}
\left [ \rho_{n, 1} h_n (x_{k j}, \hat \btheta) - \rho_{n, 1} h_n (x_{k j}, \btheta^*) \right ] \ind (\xi_p < x_{k j} \leq x). 
\label {lem:bahadur_claim_rewrite1_1}
\end{align} 
By the mean value theorem, the middle term in \eqref {lem:bahadur_claim_rewrite1_1} is 
\begin{align*} 
\rho_{n, 1} h_n (x_{k j}, \hat \btheta) - \rho_{n, 1} h_n (x_{k j}, \btheta^*)
= (\hat \btheta - \btheta^{*})^{\top} \bq (x_{k j}) 
\frac {\rho_{n, 0} \rho_{n, 1} h_n (x_{k j}, \btheta_n)} {\rho_{n, 0} + \rho_{n, 1} \exp \{ \btheta_n^{\top} \bq (x_{k j}) \}}, 
\end{align*} 
for some $ \btheta_n $ between $ \hat \btheta $ and $ \btheta^{*} $. 
Plugging it back in \eqref {lem:bahadur_claim_rewrite1_1} yields 
\begin{align} 
& \{ \hat G_1 (x) - \hat G_1 (\xi_p) \} 
- \{ G_1^{*} (x) - G_1^{*} (\xi_p) \} 
\nonumber \\ 
= & (\hat \btheta - \btheta^{*})^{\top} 
\left \{ 
n_1^{-1} \sum_{k, j} \bq (x_{k j}) 
 \frac {\rho_{n, 0} \rho_{n, 1} h_n (x_{k j}, \btheta_n)} {\rho_{n, 0} + \rho_{n, 1} \exp \{ \btheta_n^{\top} \bq (x_{k j}) \}} 
\ind (\xi_p < x_{k j} \leq x) 
\right \} 
\nonumber \\ 
= & (\hat \btheta - \btheta^{*})^{\top} O_p (n_1^{-1/4}).  
\label {lem:bahadur_claim_rewrite1_2}
\end{align} 
Here we prove the last equality. 
The expression in the curly brackets is a sum of two terms indexed by $ k = 0, 1 $. 
For the term with $ k = 0 $, because 
\[ 
\frac {\rho_{n, 0}^2 h_n (x_{k j}, \btheta_n)} {\rho_{n, 0} + \rho_{n, 1} \exp \{ \btheta_n^{\top} \bq (x_{k j}) \}} 
< \exp \{ \btheta_n^{\top} \bq (x_{k j}) \}, 
\] 
we have 
\begin{align*}
& \Big \| n_1^{-1} \sum_{j=1}^{n_0} \bq (x_{0 j}) \frac {\rho_{n, 0} \rho_{n, 1} h_n (x_{0 j}, \btheta_n)} {\rho_{n, 0} + \rho_{n, 1} \exp \{ \btheta_n^{\top} \bq (x_{0 j}) \}} \ind (\xi_p < x_{0 j} \leq x) \Big \| \\ 
= & \Big \| n_0^{-1} \sum_{j=1}^{n_0} \bq (x_{0 j}) \frac {\rho_{n,0}^2 h_n (x_{0 j}, \btheta_n)} {\rho_{n, 0} + \rho_{n, 1} \exp \{ \btheta_n^{\top} \bq (x_{0 j}) \}} \ind (\xi_p < x_{0 j} \leq x) \Big \| \\ 
= & O (1) n_0^{-1} \sum_{j=1}^{n_0} \| \bq (x_{0 j}) \| \exp \{ \btheta_n^{\top} \bq (x_{0 j}) \} \ind (\xi_p < x_{0 j} \leq x) \\ 
= & O (1) \bbE_0 [ \| \bq (X) \| \exp \{ \btheta_n^{\top} \bq (X) \} \ind (\xi_p < X \leq x)] + o_p (1), 
\end{align*} 
by the weak law of large numbers for triangular arrays. 
Note that without loss of generality, here we regard $ \btheta_n $ as a sample-independent quantity that is uniformly $ \| \btheta_n - \btheta^* \| \leq O_p (n_1^{-1/2}) $. 
Then, by the Cauchy--Schwarz inequality, uniformly over $ x $ in a $ O_p (n_1^{-1/2}) $ neighbourhood of $ \xi_p $, 
\begin{align*}
& \bbE_0 [\| \bq (X) \| \exp \{ \btheta_n^{\top} \bq (X) \} \ind (\xi_p < X \leq x)] \\ 
\leq & \sqrt{\bbE_0 \| \bq (X) \exp \{ (\btheta_n - \btheta^*)^{\top} \bq (X) \} \|^2} \sqrt{P (\xi_p < X \leq x; G_1)} \\ 
= & O (1) O_p (n_1^{-1/4}), 
\end{align*} 
where the first order assessment is uniformly over $ \| \btheta_n - \btheta^* \| \leq O_p (n_1^{-1/2}) $ by Condition~\eqref{Condition.iii} and applying the Cauchy--Schwarz inequality again; 
the second order assessment is because $ G_1 (x) $ is a continuous distribution function. 

For the term with $ k = 1 $, because 
\[ 
\frac {\rho_{n, 0} \rho_{n, 1} h_n (x_{k j}, \btheta_n)} {\rho_{n, 0} + \rho_{n, 1} \exp \{ \btheta_n^{\top} \bq (x_{k j}) \}} < 1, 
\] 
we have 
\begin{align*}
& \Big \| n_1^{-1} \sum_{j=1}^{n_1} \bq (x_{1 j}) \frac {\rho_{n, 0} \rho_{n, 1} h_n (x_{1 j}, \btheta_n)} {\rho_{n, 0} + \rho_{n, 1} \exp \{ \btheta_n^{\top} \bq (x_{1 j}) \}} \ind (\xi_p < x_{1 j} \leq x) \Big \| \\ 
= & O (1) n_1^{-1} \sum_{j=1}^{n_1} \| \bq (x_{1 j}) \| \ind (\xi_p < x_{1 j} \leq x) \\ 
= & O (1) \bbE_1 [\| \bq (X) \| \ind (\xi_p < X \leq x)] + o_p (1), 
\end{align*} 
by the weak law of large numbers for triangular arrays. 
Further, by the Cauchy--Schwarz inequality, uniformly over $ x - \xi_p = O_p (n_1^{-1/2}) $ we have 
\begin{align*}
\bbE_1 [ \| \bq (X) \| \ind (\xi_p < X \leq x)] = O_p (n_1^{-1/4}). 
\end{align*} 

Finally, by using the fact that $ \hat \btheta - \btheta^{*} = O_p (n_1^{-1/2}) $, together with \eqref {lem:bahadur_claim_rewrite1_2}, we have 
\begin{align} 
\sup_{|x - \xi_p| \leq O_p (n_1^{-1/2})} 
| \{ \hat G_1 (x) - \hat G_1 (\xi_p) \} 
- \{ G_1^{*} (x) - G_1^{*} (\xi_p) \} |
= O_p (n_1^{-3/4}). 
\label {lem:bahadur_claim_rewrite1_order}
\end{align} 

Combining \eqref {lem:bahadur_claim_rewrite1_order} and \eqref {lem:bahadur_claim_rewrite2_order}, 
the claim in \eqref {lem:bahadur_claim} is proved. 

Up to this point, the proof of Theorem~\ref {thm:bahadur} is complete. 

\end{proof}

\newpage 

\bibliographystyle{abbrvnat}
\bibliography{Archer_bib2022}

\end{document}